\documentclass[10pt]{article}
\usepackage[margin=1in]{geometry}

\usepackage{setspace}
\usepackage{authblk}
\usepackage{float}
\usepackage{amsmath}
\usepackage{amsfonts}
\usepackage{amssymb}
\usepackage{graphicx}
\usepackage{subfigure}
\usepackage{color}

\newcommand{\refSupporting}[1]{\eqref{#1}}

\newcommand{\beginsupplement}{%
		\clearpage
        \setcounter{table}{0}
       \renewcommand{\thetable}{S\arabic{table}}%
       \setcounter{figure}{0}
       \renewcommand{\thefigure}{S\arabic{figure}}%
       \setcounter{equation}{0}
       \renewcommand{\theequation}{S\arabic{equation}}%
       \setcounter{page}{1}
}
    
\newcommand{\beginmoviecaptions}{%
		\clearpage
        \setcounter{figure}{0}
        \renewcommand{\thefigure}{S\arabic{figure}}%
				\renewcommand{\figurename}{Movie}
}

\title{Balance of Mechanical Forces Drives Endothelial Gap Formation and May Facilitate Cancer and Immune-Cell Extravasation}
\author[1,2]{Jorge Escribano}
\author[2,3]{Michelle B. Chen}
\author[2,4]{Emad Moeendarbary}
\author[5]{Xuan Cao}
\author[5]{Vivek Shenoy}
\author[1,*]{Jose Manuel Garcia-Aznar}
\author[2,6,7,*]{Roger D. Kamm}
\author[6,8,*]{Fabian Spill}
\affil[1]{Department of Mechanical Engineering, University of Zaragoza, Zaragoza, Spain}
\affil[2]{Department of Biological Engineering, Massachusetts Institute of Technology, Cambridge, MA 02139, USA}
\affil[3]{Department of Bioengineering, Stanford University, Stanford, CA 94305, USA}
\affil[4]{Department of Mechanical Engineering, University College London, London WC1E 6BT, United Kingdom}
\affil[5]{Department of Materials Science and Engineering, University of Pennsylvania, Philadelphia, PA 19104, USA}
\affil[6]{Department of Mechanical Engineering, Massachusetts Institute of Technology, Cambridge, MA 02139, USA}
\affil[7]{BioSystems and Micromechanics (BioSyM), Singapore-MIT Alliance for Research and Technology, Singapore, Singapore}
\affil[8]{School of Mathematics, University of Birmingham, Birmingham, B15 2TT, UK}
\affil[*]{Correspondence and request for materials should be addressed to J.M.G.-A. (email: jmgaraz@unizar.es) or to R.D.K. (email: rdkamm@mit.edu) or to F.S. (email: f.spill@bham.ac.uk).
}

\begin{document}
	\maketitle
	
\clearpage
	
\begin{abstract}
The formation of gaps in the endothelium is a crucial process underlying both cancer and immune cell extravasation, contributing to the functioning of the immune system during infection, the unfavorable development of chronic inflammation and tumor metastasis. Here, we present a stochastic-mechanical multiscale model of an endothelial cell monolayer and show that the dynamic nature of the endothelium leads to spontaneous gap formation, even without intervention from the transmigrating cells. These gaps preferentially appear at the vertices between three endothelial cells, as opposed to the border between two cells. We quantify the frequency and lifetime of these gaps, and validate our predictions experimentally. Interestingly, we find experimentally that cancer cells also preferentially extravasate at vertices, even when they first arrest on borders. This suggests that extravasating cells, rather than initially signaling to the endothelium, might exploit the autonomously forming gaps in the endothelium to initiate transmigration.
\end{abstract}

\clearpage


\section*{Introduction}
Immune and cancer cells alike are characterized by their ability to migrate within the vasculature and then to leave the vasculature into different tissues. These processes are crucial for a functioning immune system to fight acute infections \cite{vestweber2015leukocytes} or participate in wound healing \cite{PARK2004S11}. However, chronic inflammation or tumor metastases are ultimately also initiated by extravasating immune or cancer cells, respectively \cite{grivennikov2010immunity,murray2011protective,muller2014endothelial}. Hence, while extravasation is critical to cure communicable diseases, it is also a critical contributor to virtually all non-communicable disease, ranging from cancer to asthma, atherosclerosis, rheumatoid arthritis and heart diseases \cite{libby2010inflammation,murray2011protective,mullin2005keynote}.

Much of the research on extravasation (often termed diapedesis in the context of immune cells) has focused on the role of the extravasating cell during this process, and how it interacts with the endothelial cells of the vasculature through which it is transmigrating. First, the extravasating cell needs to arrest in the vasculature. This may occur through single cells or clusters getting physically stuck in small capillaries, through the formation of adhesions, or both \cite{Reymond2013,luster2005immune,kienast2010real,wirtz2011physics,FOLLAIN201833}. Such adhesion is mediated by molecules including P-and E-selectin, ICAM, VCAM or integrins \cite{Allingham2007}. The actual process of transmigration can occur through a single endothelial cell (transcellular extravasation) or, more commonly, in between two or more endothelial cells (paracellular extravasation) \cite{martinelli2014probing,vestweber2015leukocytes}.

During paracellular extravasation, it was investigated how the extravasating cell signals to the endothelial cells, leading to weakening of VE-cadherin-mediated cell-cell junctions and subsequently gap formation, through which the cells can transmigrate \cite{Reymond2013,vestweber2015leukocytes}. Gap formation may, for instance, be stimulated by thrombin \cite{Valent2016a}. As such, molecular signaling events are firmly established as important contributors to extravasation of immune cells.

However, on a fundamental level, all the processes involved in extravasation are mechanical processes. Transmigration, like other forms of cell migration, involves the generation of mechanical forces through the actomyosin cytoskeleton \cite{CAO20161541}. Moreover, the mechanical properties of the endothelium provide passive mechanical resistance \cite{CAO20161541}. For instance, increased endothelial cell and junctional stiffness will reduce paracellular extravasation rates \cite{Schaefer4470,martinelli2014probing}. Interestingly, recent research established that active mechanical properties of the endothelial cells are also critical during endothelial gap formation \cite{liu2010mechanical,schaefer2015cell,oldenburg2014mechanical}, and the rearrangements of cytoskeletal structures are associated with changes in barrier function. For instance, a rich actin cortex parallel to cell-cell borders is associated with stabilized VE-cadherin junctions and thus tight barriers \cite{ando2013rap1,dorland2017cell}, whereas actomyosin stress fibers pulling radially on junctions can lead to junctional remodeling \cite{liu2010mechanical,Huveneers2012}. Additionally, actin-rich pores can actively contract to prevent leakage during extravasation \cite{heemskerk2016f}. However, there is still a lack in mechanistic and systems level understanding of the different roles of active and passive mechanical properties of the endothelium.

Mathematical multiscale models are powerful tools to investigate the interplay of different physical drivers of biological processes. Many different approaches have been employed to model and understand the dynamics of epithelial monolayers. Agent based models, where individual cells are explicitly taken into account, include center based models (CBM)\cite{Galle2005}, vertex models \cite{fletcher2014vertex,Lin2016} and deformable models (DFM)\cite{Jamali2010, Pathak2016}. However, these models do not explicitly model cell-cell adhesion dynamics in a way that leads to the experimentally observed gap formation in monolayers of endothelial cells, and can thus not easily be employed to study this problem so crucial for cancer and immune transmigration.

In this paper, we introduce a mathematical multiscale model of the mechanics of an endothelial monolayer where each endothelial cell contains contractile actin structures that may contract radially or in parallel to the plasma membrane. Then, cells are tethered to neighboring cells by cell-cell junctions that can dynamically form and break in a force-dependent manner. We employ this model to investigate the mechanisms of gap formation in an endothelial monolayer. Interestingly, we find that gaps open dynamically in the absence of any extravasating cells. These gaps form preferentially at the vertices where three or more endothelial cells meet, as opposed to the borders in between two cells. This is in line with our experimental data obtained in-vitro from quantifying gap formation of monolayers of {\color{black}human umbilical vein endothelial cells} (HUVECs) seeded on glass. We quantify the frequency of gap openings as well as the duration of gap openings, obtaining good agreement between numerical predictions and experiments. Moreover, through multi-dimensional parameter studies, the mathematical model is able to give us insights into the physical and molecular drivers of the gap formation and gap dynamics. The model predicts that active and passive mechanical forces play an important role in the initial gap formation and in controlling size and lifetime of gaps once they initially formed. The catch bond nature of the cell-cell adhesion complexes as well as the force-dependent reinforcement of adhesion clusters may both stabilize junctions in response to forces acting on them. However, while the catch bonds ultimately weaken when forces are increased beyond the maximal lifetime of a single molecular bond, the force-dependent reinforcement will increase adhesion strength with increasing force \cite{Panorchan2006,liu2010mechanical}. While the catch bond nature and the force dependence of the adhesion clustering processes both crucially influence gap opening frequencies, we find that gap lifetime and gap size are even more sensitive to the passive mechanical properties of the cell. Increased stiffness of the membrane/cortex and, even more notably, of the actin stress fibers will reduce lifetime and size, since the cells will then increasingly resist opening gaps through counteracting forces. On the other hand, we find that {\color{black}changes} in bending stiffness of the membrane/cortex may have gap promoting or inhibiting effects. 

Our model predictions of gap opening frequency and lifetime at both cell vertices and borders are validated by experiments observing such gaps in endothelial monolayers in the absence of any extravasating cell. The results thus challenge the paradigm that all extravasating cells primarily cause gap opening through interactions with the endothelium \cite{vestweber2015leukocytes,Reymond2013,Yeh18122017}. We then show experimentally that extravasating cancer cells indeed primarily extravasate at vertices, in line with similar observations for neutrophils \cite{burns1997neutrophil}. Moreover, we show that cancer cells prefer to extravasate at vertices even when they initially attached to the endothelium at two-cell borders. {\color{black} This suggests that, even though extravasating cells can actively interact with the endothelium during transmigration, as shown in earlier studies, they may also take advantage of the autonomous occurrence of a gap, as predicted and verified to occur in our model}. 
In summary, our work highlights the importance of taking the dynamic and autonomous mechanical properties of the endothelium into account when trying to understand gap formation and extravasation.

\section*{Computational Model of Endothelial Monolayers}
We present a novel model of an endothelial cell (EC) monolayer that incorporates different intracellular mechanical structures and dynamical cell-cell adhesions. The intracellular mechanical state is determined by radial contractile actin stress fibers and the cell membrane together with the actin cortex. For simplicity, we combined membrane and cortex into single viscoelastic elements, composed of an elastic spring and a viscous damper, that we refer to, from now on, as membrane elements. The radial stress fibers are also modeled by viscoelastic elements with different mechanical properties from the membrane, similar to a model of epithelial cells \cite{Jamali2010} (see Fig. \ref{fig:vertexVersusEdgeGaps}A). 
Neighboring cells may form cell-cell adhesions at adjacent nodes, and the resulting adhesion bond is modeled through a spring. The passive mechanical properties of the monolayer are thus modeled through a network of connected elastic and viscoelastic elements, similar to models of epithelial sheets \cite{Pathak2016,Jamali2010}. 
Since we are interested in studying the opening dynamics of gaps in the endothelial barrier, we explicitly simulate the dynamical binding of adhesion complexes. Contractions represent myosin motor activity that is known to exhibit randomness \cite{guo2014probing}, so we employ Monte-Carlo simulations to estimate the occurrence of such forces as well as that of protrusive forces due to actin polymerization. The forces are then redistributed across the network of connected viscoelastic elements. Cell-cell adhesion complexes that mechanically link neighboring cells can dynamically bind and unbind in a force-dependent manner. The adhesion complexes in the model provide an effective description of both bonds of cell-cell adhesion molecules (such as VE-cadherin) and bonds of these adhesion molecules to the cytoskeleton. Cadherins and adhesion-cytoskeleton bonds are known to increase their binding strength in response to smaller forces, before they ultimately rupture \cite{Lecuit2015}. This catch-bond type behavior is included in our model, and unbinding is thus simulated through a force-dependent Monte-Carlo simulation. Moreover, the number of VE-cadherins in an adhesion complex is modeled through a force-dependent adhesion clustering mechanism, as described in \cite{liu2010mechanical, Huveneers2012, Chen2016a, Barry2015, Yonemura2010}. A more detailed description of the mathematical model and its numerical implementation is given in the supplementary information (SI).

We employ our endothelial monolayer model to explore the dynamics of endothelial cell junctions. We predict the frequency, size and duration of gaps, as well as the preferred geometrical locations of the gap formation, and compare the predictions with our experimental measurements. The parameters used in the simulations are detailed in Table \ref{tab:1}. After comparing our predictions with the experimental results, we perform sensitivity analyses to investigate how cell mechanical properties, cell-cell adhesion characteristics and myosin generated forces regulate the formation, lifetime and size of gaps in the endothelium.

{\color{black}
\subsection*{Summary of major model parameters}
Here we present a summary of the major parameters of the model that had a significant impact on our model behavior, and were consequently thoroughly investigated through sensitivity analysis in the remainder of this paper. Table \ref{tab:small} lists all these parameters, and for a complete list and discussion see the Supporting Information. The main parameters investigated are related to cell mechanical properties, adhesion properties or myosin force generated processes.

Cell mechanical properties are dictated by stress fiber stiffness ($K_{sf}$), membrane stiffness ($K_{memb}$) and bending stiffness (incorporated through a rotational spring constant, $K_{bend}$). Stress fiber stiffness controls the rigidity of the interior of the cell, whereas membrane stiffness controls the rigidity of the membrane and the adjacent actin cortex. Bending stiffness acts on the membrane nodes depending on the relative orientation between the edges connecting at a given node. 

Adhesion properties are controlled by the mechanical properties of the adhesion complexes and their binding and unbinding rates. Adhesion complex mechanics are modeled by linear springs, controlled by their stiffness constant, $K_{adh}^{0}$. The binding rate depends on distance and can be controlled by the adhesion complex density, $\rho_{adh}$. We then model the reinforcement of a bond that is already formed by the additional recruitment of adhesive proteins into the bond. Reinforcement is force dependent and can be controlled by the binding rate constant for adhesion reinforcement, $k_{reinf}^{0}$. Unbinding follows a catch bond behavior. The catch bond unbinding curve can be modified through two rate coefficients: $k_{s}^{0}$, which represents a slip bond, and $k_{c}^{0}$, which is the additional parameter characterizing the initial increase in the bond lifetime with force (see Fig. \ref{SI:fig:CatchSlipExplan}).

Then, the model includes contractile forces due to myosin motor activity, and protrusive forces that may arise due to actin polymerization. These forces can be directed radially (following the stress fibers direction) or in a tangential direction (following membrane segments). In the sensitivity analysis we have varied the magnitude of contraction forces in the radial direction ($F_{Radial}$) and in the tangential direction ($F_{cortex}$).

\begin{table*}[!ht]
	{\color{black}	       
		\begin{tabular}{ll}
			\hline\noalign{\smallskip}
			Parameter& Symbol\\
			\noalign{\smallskip}\hline\noalign{\smallskip}
			Stress fiber stiffness & $K_{sf}$ \\
            Membrane stiffness & $K_{memb}$ \\
			Rotational spring constant & $K_{bend}$  \\
            Adhesion complex stiffness constant per bond& $K_{adh}^{0}$ \\
            Adhesion complex density	&  $\rho_{adh}$  \\
            Binding rate for adhesion reinforcement constant& $k_{reinf}^{0}$ \\ 
            Unbinding rate coefficient for catch curve& $k_{c}^{0}$\\
            Unbinding rate coefficient for slip curve& $k_{s}^{0}$ \\  	
			Maximum force due to radial contraction & $F_{Radial}$ \\
			Maximum force due to cortical tension & $F_{cortex}$ \\		
		\end{tabular}
		\caption{List of parameters used in the sensitivity analysis.}
		\label{tab:small}
        }
\end{table*}

}
\section*{Results}	\label{sec4}

\subsection*{Gaps open preferentially at vertices}
\begin{figure}[!ht]
	\centering
	\includegraphics[width=0.98\linewidth]{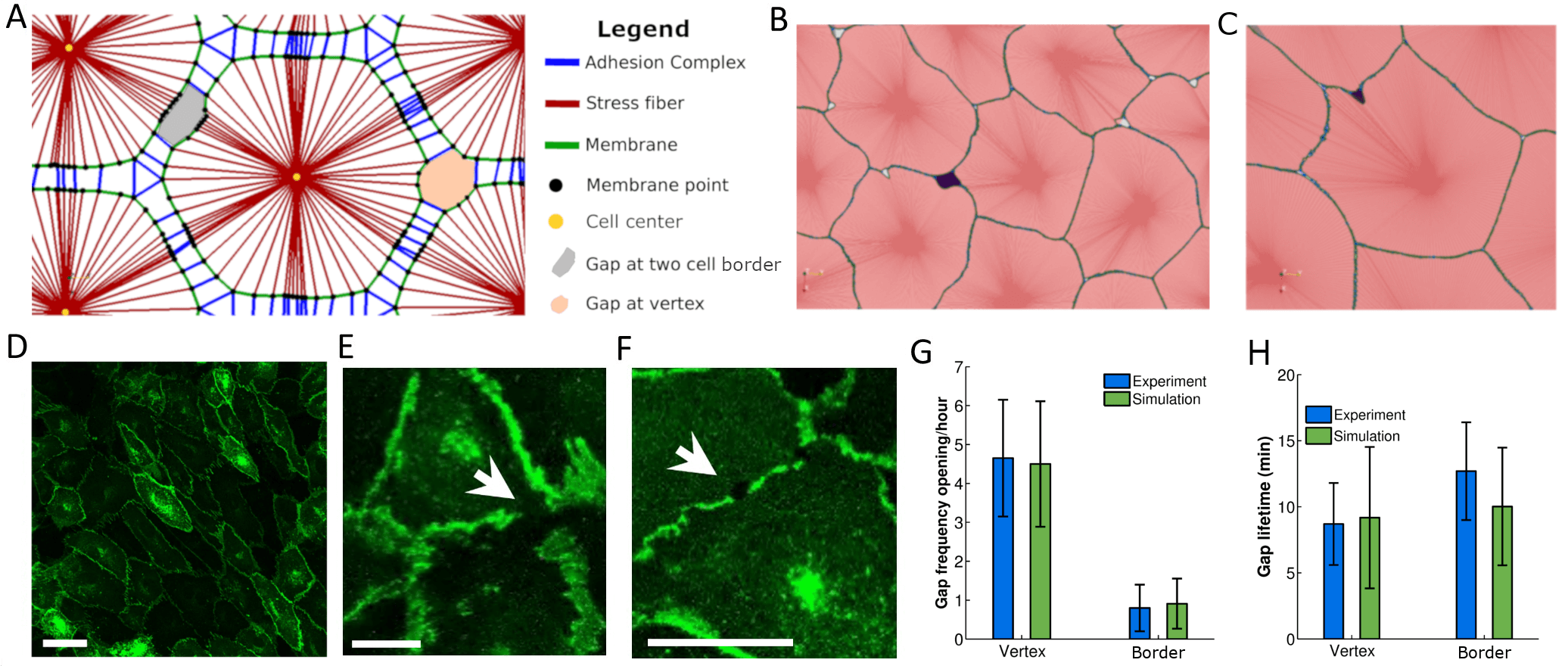}
	\caption{{\bf Endothelial Gaps open preferentially at vertices.} A: Main components of the model. Cells are formed by stress fibers (red), membrane segments (green) and membrane points (black). Membrane points are binding sites at which cell-cell adhesion complexes (blue) may connect to membrane points on adjacent cells and thus mechanically link these cells. When these adhesions  break as a consequence of the force-dependent binding law, gaps in the endothelium are generated. Gaps can be generated at a two cell border (grey) or at a vertex between three or more cells (orange). B, C: Simulation of endothelial monolayer dynamics. Green denotes cell membrane and red the inside of a cell (darker red are the stress fibers that compose the cell ). Dark purple denotes a detected gap. B: Gap at a vertex. C: Gap at border. D-F: Endothelial monolayer of HUVEC cells expressing VE-cadherin-GFP on glass. Gap opened at vertex (E) and border (F). Scale bars are $50\mu m$ (D) and $20\mu m$ (E,F), respectively. G, H: Quantification of gap opening frequency and gap lifetime at vertices or borders, respectively. {\color{black} Simulations correspond to the reference case. Error bars show the standard deviation.}}\label{fig:vertexVersusEdgeGaps}
\end{figure}

\begin{figure*}[ht]
		\centering
		\includegraphics[width=0.95\linewidth]{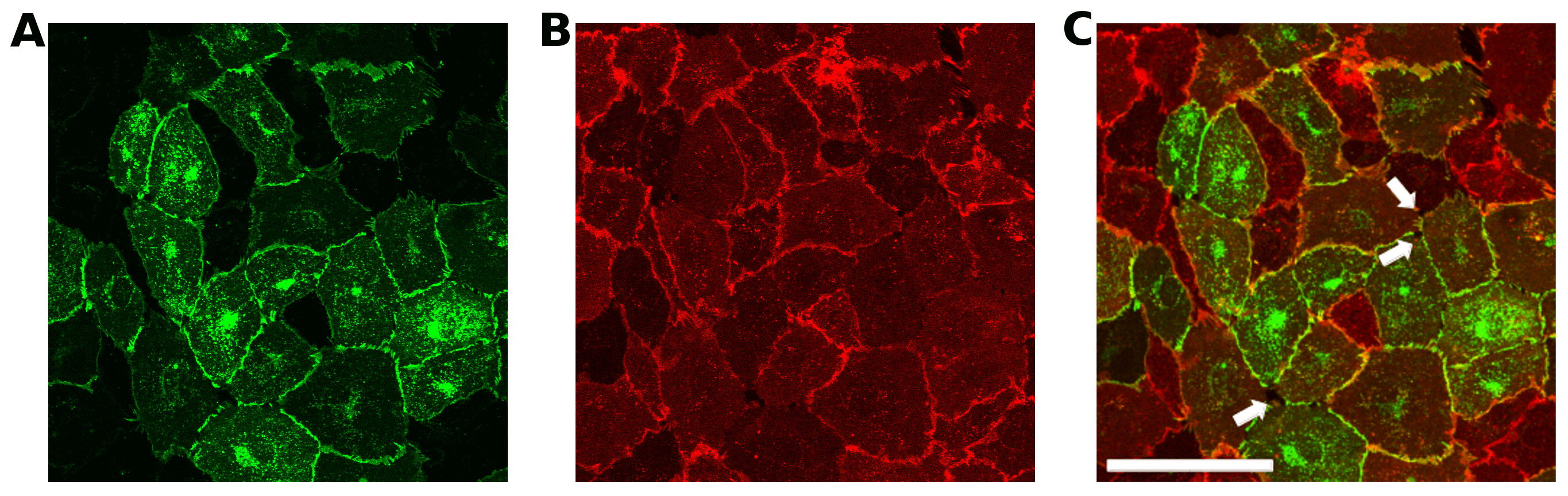}
		\caption{\textbf{Gaps in VE-cadherin correspond to gaps in CD31.} Endothelial monolayer stained with VE-cadherin (green, A) and CD31 (red, B). C: Merged image confirms that gaps observed within the VE-cadherin mediated cell-cell adhesions are also present within CD31, indicating that gaps seen in VE-cadherin are real physical gaps between the cells. Scale bar $100\mu m$.
        }\label{SI:fig:VEcadCD31Control}
\end{figure*}

\begin{figure*}[ht]
		\centering
		\includegraphics[width=0.8\linewidth]{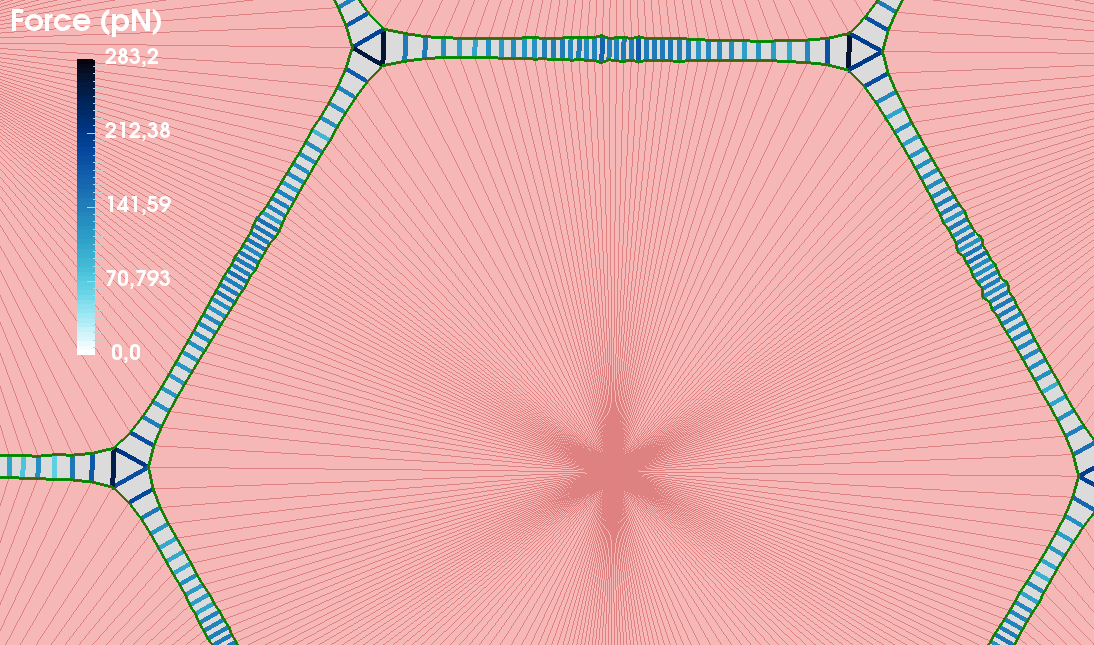}
		\caption{\textbf{Stresses on the cell-cell adhesions.} Homogeneous contractions are applied to all the hexagonal cells in the monolayer. Stresses concentrate on the adhesions at vertices, as opposed to the adhesions at border.}\label{SI:fig:stress}
\end{figure*}

Fig. \ref{fig:vertexVersusEdgeGaps}B,C and Movie \ref{SI:mov:ReferenceCase} show typical simulations of the monolayer dynamics of the computational model. We observe that gaps open preferentially at vertices, i.e. the intersections of three or more cells, as opposed to the border between two cells. We have quantified this by counting the total number of gaps formed as well as their lifetime at borders and vertices of the cell in the center of the monolayer, and showed that our model predictions are in line with the experimental observations (Fig. \ref{fig:vertexVersusEdgeGaps}G,H). {\color{black} These experiments were performed by seeding HUVEC cells on glass, where they formed a continuous monolayer. The gaps were experimentally quantified through inspection of visible gaps within the VE-cadherin-GFP signal in the monolayer (arrows in Figs. \ref{fig:vertexVersusEdgeGaps}E,F). Controls simultaneously showing VE-cadherin-GFP and CD31 staining show that the VE-cadherin gaps are also visible in the CD31 staining, indicating that the VE-cadherin gaps correspond to real physical gaps between two or more cells \ref{SI:fig:VEcadCD31Control} (see Methods for further details of the experimental setup and quantification)}. Vertices are points where more than two cells exert forces and where tangential force components naturally propagate to. Therefore, it is expected that stress concentrates at the three cell vertex rather than at the two cells borders, and the simulations confirm this hypothesis (Supplementary Fig. \ref{SI:fig:stress} and Movie \ref{SI:mov:stress}). The forces on adhesion clusters at the vertices are thus more likely to exceed the corresponding force of maximal lifetime of the bonds, as will be discussed in more detail below.

\subsection*{Mechanical properties of cell-cell adhesion complexes limit endothelial gap opening frequency}
	
\begin{figure}[!ht]
	\centering		
    \includegraphics[width=0.96\linewidth]{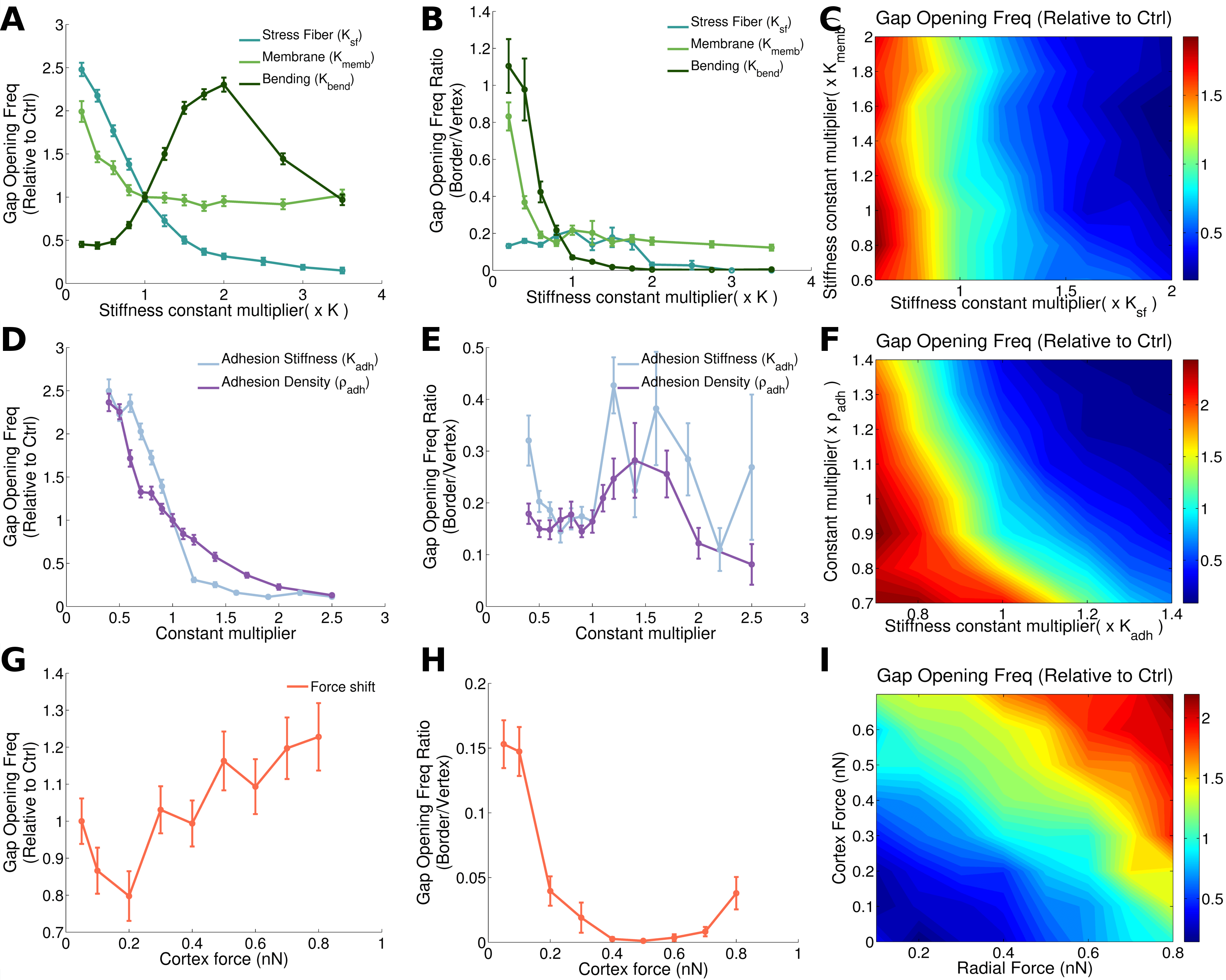}
	\caption{\textbf{Intracellular and cell-cell junctional mechanical properties dictate gap opening dynamics.} The first column (A,D,G) correspond to the total number of gaps (vertex plus border) generated per time, compared to the reference case {\color{black} (which, by definition, corresponds to the value $1$ on the y-axis)}. The second column (B,E,H) shows the ratio of gaps that occur at a two cell border to the gaps that originate at a three cell vertex. {\color{black} Error bars show to the standard error.} The third column (C,F,I) shows the impact of a two parameter variation in the gap opening frequency. The first row shows results from varying cell mechanical properties (stress fiber, membrane and bending stiffness (A,B)), and simultaneous variations of stress fiber and membrane stiffness (C). In the second row (C-F) properties of cell-cell junction are changed: adhesion stiffness and adhesion density. The third row shows results for increasing cortical force, keeping the total force constant (G and H) or varying both forces simultaneously (I).}\label{fig:parameterImpact}
\end{figure}

	We study how variations in the mechanical properties of the cells, the cell-cell adhesion complexes or force variations affect the rate of gap formation. Fig. \ref{fig:parameterImpact}A,B show how passive mechanical properties of the cell affect both the frequency (Fig. \ref{fig:parameterImpact}A) and the location of the gap openings (Fig. \ref{fig:parameterImpact}B). Increasing stiffness of either the membrane or the stress fibers provokes a decrement of the gap generation frequency (Fig. \ref{fig:parameterImpact}A and Movies \ref{SI:mov:lowKsf},\ref{SI:mov:highKsf}). This is intuitive, since increasing stiffness stabilizes the movements of cells and makes the monolayer less dynamic. On the other hand, the location of the gap openings (i.e. whether they occur at a vertex or border) is critically affected by membrane stiffness at low values, until it stabilizes for intermediate and high membrane stiffness. In contrast, stress fiber stiffness affects gap location for very high stiffness, where gaps are almost fully prevented from opening at the borders (Fig. \ref{fig:parameterImpact}B). 
    Interestingly, increasing bending stiffness first increases gap generation up to a maximum point, before it leads to a decrease in gap opening frequency (Fig. \ref{fig:parameterImpact}A). For small to intermediate bending stiffness, the frequency of gap openings increases, since bending stiffness is critical for effective force propagation between neighboring adhesion sites at vertices. When a single adhesion complex ruptures, bending stiffness leads to increased forces on neighboring adhesion complexes. After a peak in gap opening frequency at intermediate bending stiffness, a drop in the gap formation is observed for higher bending stiffness. This is caused by the resulting stabilization of the existing gaps at vertices. This high bending stiffness opposes sharp corners of the membrane at vertices and thus favors stable gaps that are permanently open, implying no new gaps are formed (Movie \ref{SI:mov:highKbend}). On the other hand, at cell borders, a high bending stiffness implies that if a single adhesion cluster is ruptured, the forces on it are redistributed across many neighboring adhesion sites and this stabilizes the borders (Fig. \ref{fig:parameterImpact}B).
    
Turning to the role of cell-cell adhesion complex properties, our model shows that as the junctions become more stable, gaps open less frequently (Fig. \ref{fig:parameterImpact}D). To increase cell-cell junction stability, we increase the mechanical stiffness of individual adhesion bonds, or the density of adhesion molecules. These results are in line with previous experimental work \cite{martinelli2014probing}, which reported that more stable cell-cell junctions result in fewer transmigrating cells. While the total number of gaps at either vertex or border decreases with increasing cell-cell adhesion complex stiffness or cell-cell adhesion density available for binding, we see that there are no significant differences between gaps generated at the vertex and gaps generated at the borders (Fig. \ref{fig:parameterImpact}E). 

Fig. \ref{fig:parameterImpact}G,H show the impact of {\color{black}changing} the cortical and radial forces, where the total force is kept constant (when the radial force decreases, the cortical force is increased by the same magnitude). This is biologically relevant since cells are known to shift their cytoskeletal compartments in a context dependent manner \cite{suarez2016internetworkActinCompetition}. In fact, cell monolayers subjected to shear flow have been reported to increase cortical actin while decreasing stress fibers \cite{martinelli2014probing}. Endothelial cells in particular, are known to exhibit both radial and tangential stress fibers with a different effect on gap opening dynamics \cite{Oldenburg2014}. As the force shifts from radial to cortical forces, total gap formation fluctuates with a slight increase as cortical forces increase (Fig. \ref{fig:parameterImpact}G). For high cortical forces, the gaps also clearly tend to localize more at the vertices (Fig. \ref{fig:parameterImpact}H). This is because contractions parallel to the membrane result in force concentrations at the vertices. For very high cortex forces, the typical stresses on adhesion clusters at the vertices  may thus be higher than the force where the lifetime of catch bonds peaks (Supplementary Fig. \ref{SI:fig:CatchSlipExplan}), explaining the small increase in the number of gaps formed (Fig. \ref{fig:parameterImpact}G). On the other hand, we will later show that these gaps formed at high cortical forces are typically small and have a short lifetime, limiting their potential for extravasation (see Fig. \ref{fig:SensLifeSize}I,J). 
	
	To take into account that molecular or physical  perturbations may simultaneously affect multiple parameters, we now study how variations of pairs of these parameters at the same time may influence the monolayer integrity and the localization of the gap formation. Although, we have previously seen in Fig. \ref{fig:parameterImpact}A that membrane and stress fiber stiffness have a similar effect on the gap opening frequency, in Fig. \ref{fig:parameterImpact}C we can observe how the effect of varying stress fiber stiffness is clearly predominant over the effect of varying membrane stiffness. 
      Fig. \ref{fig:parameterImpact}F shows the impact of varying cell-cell adhesion stiffness and cell-cell adhesion complex density available for binding. Interestingly, there is a synergy between both parameters on regulating gap opening frequency, as evident through the curved shape of the levels of equal gap opening frequency (Fig. \ref{fig:parameterImpact}F).			
	In Fig. \ref{fig:parameterImpact}I we show the combined role of cortex and radial forces, thus not keeping total force fixed as in Fig. \ref{fig:parameterImpact}G,H. This confirms that total force is the main driver of gap opening frequency, as opposed to a redistribution of forces between cortex and stress fibers (Fig. \ref{fig:parameterImpact}I).

\subsection*{Passive cell-mechanical properties dictate endothelial gap lifetime and size}
\begin{figure}[!ht]
		\centering
		\includegraphics[width=0.98\linewidth]{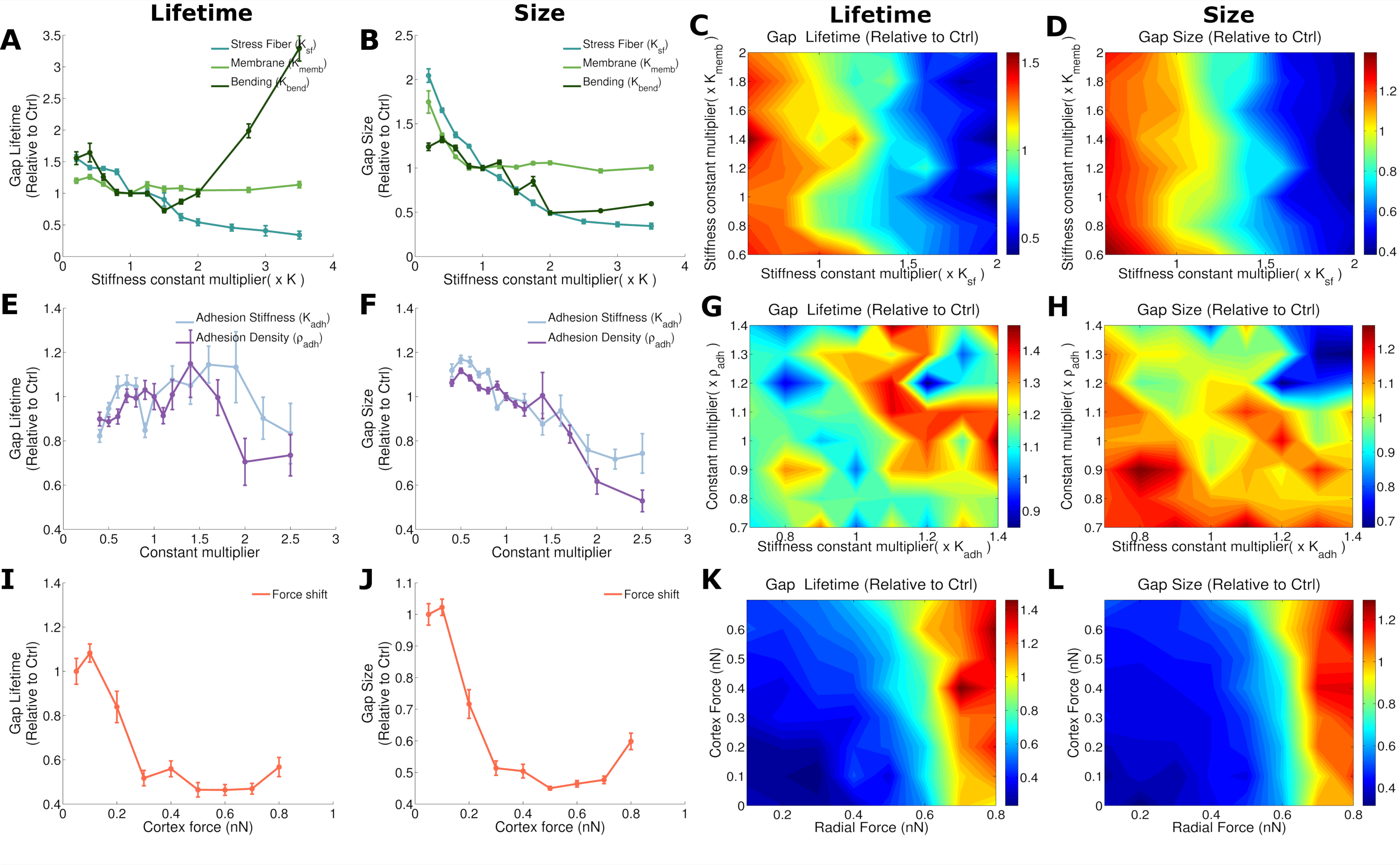}
		\caption{\textbf{Impact of Cell mechanics and cell-cell junctions properties on lifetime and size of gaps.} Average lifetime of gaps ratio (divided by control case) in first column and average size in second column. {\color{black} Note that the point where the x and y axes values are 1 corresponds to the reference case.} {\color{black} Error bars show to the standard error.} A,B: Impact of membrane, stress fiber and bending stiffness. E,F: Impact of cadherin properties: stiffness and cadherin density (which affects the binding probability). I,J: Changing cortical force while keeping total force constant. Effect of two parameter variation on gap lifetime (third column) and size (fourth column). C,D: variation of membrane and stress fiber stiffness. G,H: Adhesion stiffness versus {\color{black}adhesion complex density variation.} K,L; Cortical and radial force variation.
		}\label{fig:SensLifeSize}
\end{figure}

The lifetime and size of a gap are physical parameters that may also limit a cancer or immune cell's potential to extravasate through the monolayer. Here, we show how the lifetime and size of a gap are influenced by cell mechanical and junction properties, without the presence of extravasating cells (Fig. \ref{fig:SensLifeSize}). We observe that membrane stiffness has a marginal influence on the life time of the gap, whereas increasing stress fiber stiffness clearly reduces the time that a gap is open and the gap size (Fig. \ref{fig:SensLifeSize}A,B). Indeed, higher stress fiber stiffness will result in mechanical resistance to an opening gap and thus inhibit the propagation of the defect in the cell-cell junctions, leading to a stabilization of the monolayer (see Movies \ref{SI:mov:lowKsf}, \ref{SI:mov:highKsf}). The dominance of stress fiber stiffness over membrane stiffness in regulating lifetime and size remains valid in a broad range of parameter values (Fig. \ref{fig:SensLifeSize}C,D). 

Interestingly, increasing bending stiffness to high values may increase gap lifetime (Fig. \ref{fig:SensLifeSize}A). This is because higher bending stiffness will resist deviations from straight membranes. Thus, at straight borders, higher bending stiffness will resist gap openings whereas at vertices with high curvature, cells are more likely to adapt their shape resisting high curvature, thus favoring opened gaps. The dynamics of the monolayer for low bending stiffness is shown in Movie \ref{SI:mov:lowKbend}.

Fig. \ref{fig:SensLifeSize}E,G show that adhesion complex stiffness and density at low values do not have a big impact on lifetime, however as they increase, lifetime starts to decrease. Both stiffness and density have a similar effect, since the total stiffness of an adhesion complex depends on both density and single bond stiffness (Eq. \refSupporting{eq:adhesionStiffness}). Higher stiffness of the adhesion complex leads to more passive mechanical resistance to gap openings, and this effect dominates for high stiffness. The level of noise due to repeats of our MC simulations is higher for these adhesion parameters than for the parameters determining cell mechanics. Likewise, for the gap size, the stabilizing effect of both adhesion complex stiffness and density dominates and leads to a reduction in gap size (Fig. \ref{fig:SensLifeSize}F,H). However, the effect of increasing the density is slightly stronger than that of increasing single bond stiffness. This is because the density affects not only adhesion complex stiffness (Eq. \refSupporting{eq:adhesionStiffness}), but also the rate of forming new adhesion complexes (Eq. \refSupporting{eq::bind}) and the rate of reinforcing existing bonds (Eq. \refSupporting{eq:reinforcement}). These effects together thus synergize to stabilize gaps and prevent them from growing too large.

Earlier, we have shown that a shift in the force (from radial to cortical) produces an increment in gap formation (Fig. \ref{fig:parameterImpact}G). Fig. \ref{fig:SensLifeSize}I,J show that this shift in the force reduces gap lifetime and size. This indicates that, although the frequency of opening is increased, these gaps are smaller and last shorter in time which may reduce paracellular extravasation, as suggested in previous experimental work \cite{martinelli2014probing}. Combined changes of cortical and radial force show that although both kinds of forces are needed to increases gap size and lifetime, the impact of radial forces is clearly predominant over the impact of cortex forces (Fig. \ref{fig:SensLifeSize}K,L). This is intuitive, since radial forces clearly separate cell borders generating bigger gaps and make them harder to close, whereas cortical forces distribute forces to vertex regions. This does not provoke large cell deformations, which is reflected in the low impact on the gap size and lifetime observed.
	
\subsection*{Catch bonds facilitate regimes of maximal endothelial stability}
\begin{figure}[!ht]
\centering
\includegraphics[width=0.98\linewidth]{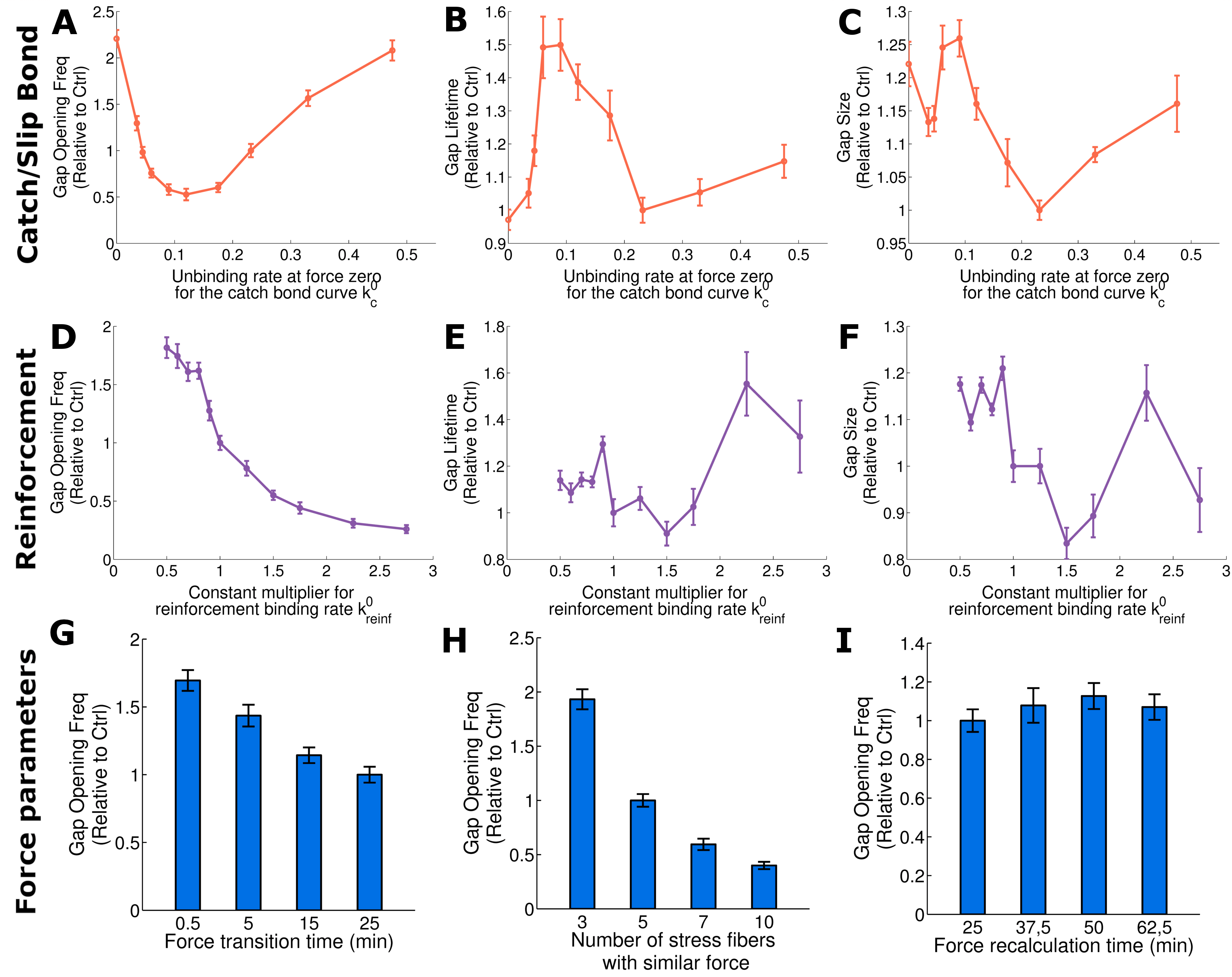}
\caption{\textbf{ Effect of the maximal lifetime of a catch bond, the cadherin reinforcement and the force application on the gap opening dynamics.} First row (A-C) shows the impact of shifting from a pure slip bond ($k_c^0 = 0 $) to a catch bond. As $k_c^0$ increases, the peak of stability moves to higher force while we fix the magnitude of a single bond lifetime. Second row  (D-F) shows reinforcement analysis varying $k_{reinf}^{0}$. Results are normalized with reference case values and are shown for gap opening frequency (A, D), gap lifetime (B, E) and gap size (C, F). Third row shows the effect of force application on total gap opening frequency. G: Total gap opening frequency depending on the time that takes to make the force transition. Longer time means smoother force changes. H: Total gap opening frequency depending on the number of stress fibers at which the same force is distributed. I: Variation in force recalculation time for all types of forces considered in the model. {\color{black} Note that the reference case corresponds to the point where y and x axes values are 1 in figs A to F.  Error bars show the standard error.}}\label{fig:catch_slip_force}
\end{figure}

In Fig. \ref{fig:catch_slip_force}A-C we show the impact of varying the catch-bond unbinding parameter $k_c^0$ that shifts the location of the peak of maximal lifetime of a single catch bond, while we maintain the actual maximum value through simultaneously shifting the slip-bond unbinding parameter $k_s^0$ (Eq. \refSupporting{eq:catchbondunbinding} and Fig. \ref{SI:fig:CatchSlipExplan}). We observe that for a pure slip bond (corresponding to $k_c^0=0$), gaps occur at a higher rates than for small nonzero values of $k_c^0$. Increasing $k_c^0$ further leads to a minimum in gap opening frequency, from which the frequency increases again. This minimum corresponds to a maximum of stability, where forces on the adhesion complexes are similar in magnitude to the peak of stability of the catch bond. Consequently, shifting the location of that peak even further towards higher forces (by increasing $k_c^0$ even further) means we destabilize the catch bonds again. Note that the gap lifetime and size of gaps are much less influenced by the location of the catch bond maximum than the gap opening frequency.

 In Supplementary Fig. \ref{SI:fig:CatchSlipHist}, we show histograms of the forces on adhesions comparing the number of bound clutches, the number of unbinding events, and the ratio of unbound to total bonds for slip bonds ($k_c^0=0$) to the catch bond with reference values ($k_c^0=0.27 s^{-1}$). We see that adhesions in the catch bond case bear and disengage at higher forces than for the slip bond case, confirming that the typical forces on bonds are of such magnitude that the catch bond nature stabilizes the junctions. Obviously, shifting to even higher values of $k_c^0$ would result in the unbinding of most bonds (not shown).

In Fig. \ref{fig:catch_slip_force}D-F we modify the reinforcement binding rate $k_{reinf}^{0}$ to check the influence of the reinforcement. This is different from the previous analysis where the adhesion complex density available for binding was changed, since now the binding probability based on distance is not affected (Eq. \refSupporting{eq::bind}). However, we see the same trend of increasing stability with increasing $k_{reinf}^{0}$ (Fig. \ref{fig:catch_slip_force}D), in line with the result obtained from varying cadherin density (Fig. \ref{fig:parameterImpact}D), suggesting that binding is mainly regulated by this reinforcement process. Similar to the catch bond, we see that adhesion reinforcement is less important in determining gap size or lifetime (Fig. \ref{fig:catch_slip_force}E,F) than in regulating gap opening frequency.

\subsection*{Force fluctuations and distribution regulate gap opening dynamics}
We have shown that both the magnitude of forces and the cytoskeletal compartment that generates the forces (stress fibers or cortex) affect gap opening frequency, size and and lifetime. Besides these broad compartments, many other biological and physical parameters affect how forces ultimately act on cell-cell junctions: Forces may act in a directed manner due to larger parallel actin bundles and synchronous myosin activation, e.g. initiated through waves of activators \cite{Valent2016a}, or may act more randomly \cite{guo2014probing}. We test variations in force applications through parameters that affect the transition time when forces change ($t^{Force}_{Transition}$), through spatial force distributions and through the velocity at which forces are modified. In Fig. \ref{fig:catch_slip_force}G we observe how increasing the force transition time $t^{Force}_{Transition}$ slowly reduces the gap opening. This is due to the fact that a slower, persistent application of forces leads to a redistribution of the forces through rearrangement and remodeling of the cell. It is consistent with experimental works that showed that force fluctuations influence gap opening dynamics \cite{Valent2016a}. 

\begin{figure*}[ht]
	\centering
	\includegraphics[width=0.8\linewidth]{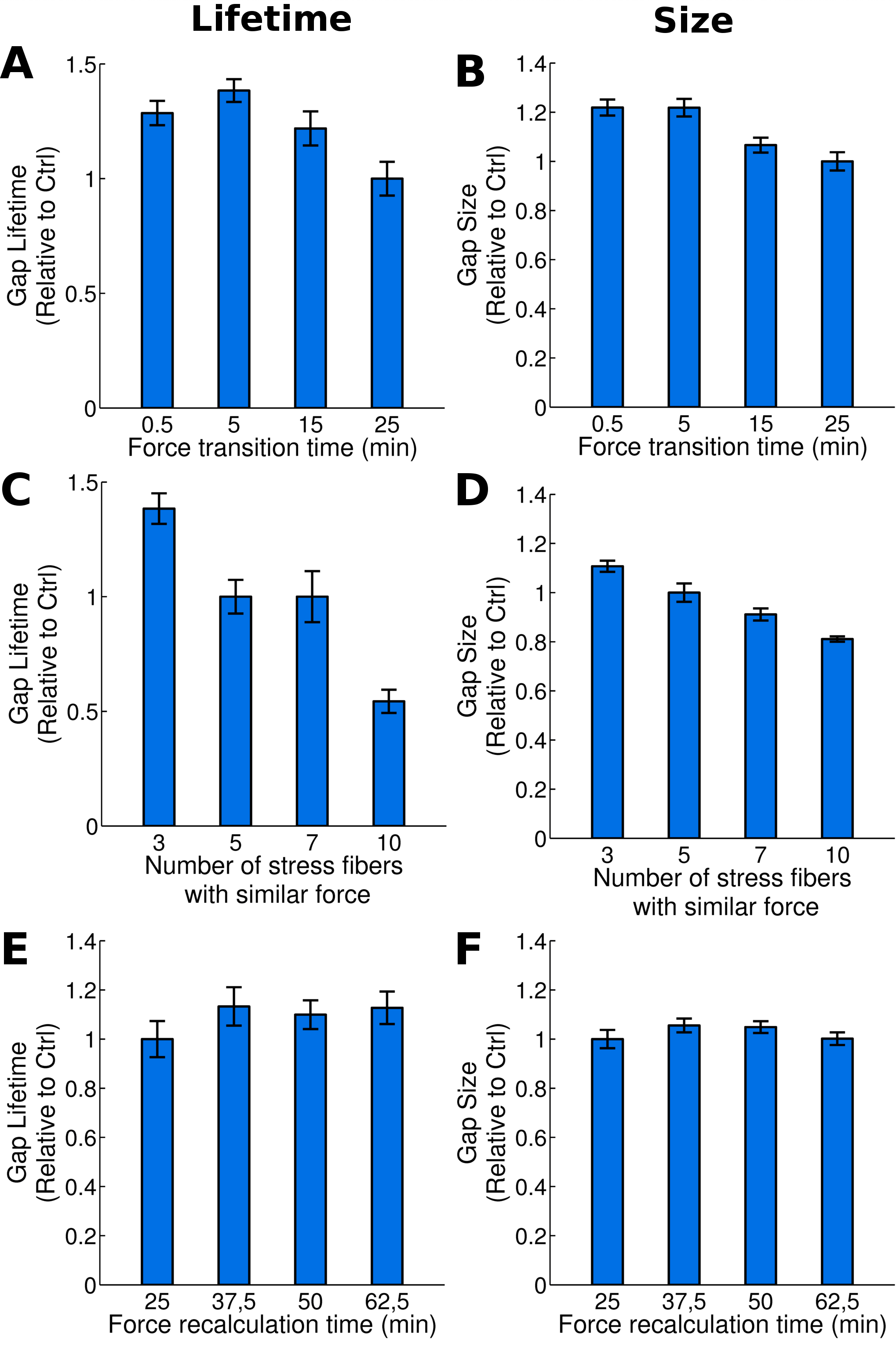}
	\caption{\textbf{Effect of force application on gap size and lifetime.} Corresponds to Fig. \ref{fig:catch_slip_force}. Left column corresponds to lifetime and right column to size. (A, B): Changes in the transition time of the application of the recalculated forces. Longer time means smoother force changes. (C, D) Variation in the number of stress fibers over which the same force is distributed. (E, F) Variation in force fluctuation time for all types of forces considered in the model. {\color{black} Error bars show to the standard error.}}\label{SI:fig:force_appLifeSize}
\end{figure*}

Then, distributing the same radial forces over several adjacent stress fibers reduces gap opening frequency (Fig. \ref{fig:catch_slip_force}H). More spatially distributed forces are less capable of damaging cell-cell junctions than localized peak forces, since such high peak forces are required to overcome the catch bond maximal lifetime. Likewise, high peak forces lead to longer lifetime and larger size of the resulting gaps (Supplementary Fig. \ref{SI:fig:force_appLifeSize}C,D).

Next we observe the effect of force persistence in time. We vary the force recalculation time parameter (equally for all forces) in Fig. \ref{fig:catch_slip_force}I. Results show that the time that forces are applied does not have a big influence on gap formation. This suggests that cells are able to adapt to forces in longer time scales and therefore it is not the time that forces are applied what regulates gap formation, but the transitions of force fluctuations and their spatial distribution.

\subsection*{Cancer extravasation mimics autonomously occurring endothelial gap dynamics}
	
\begin{figure}[h!]
\centering
\includegraphics[width=0.98\linewidth]{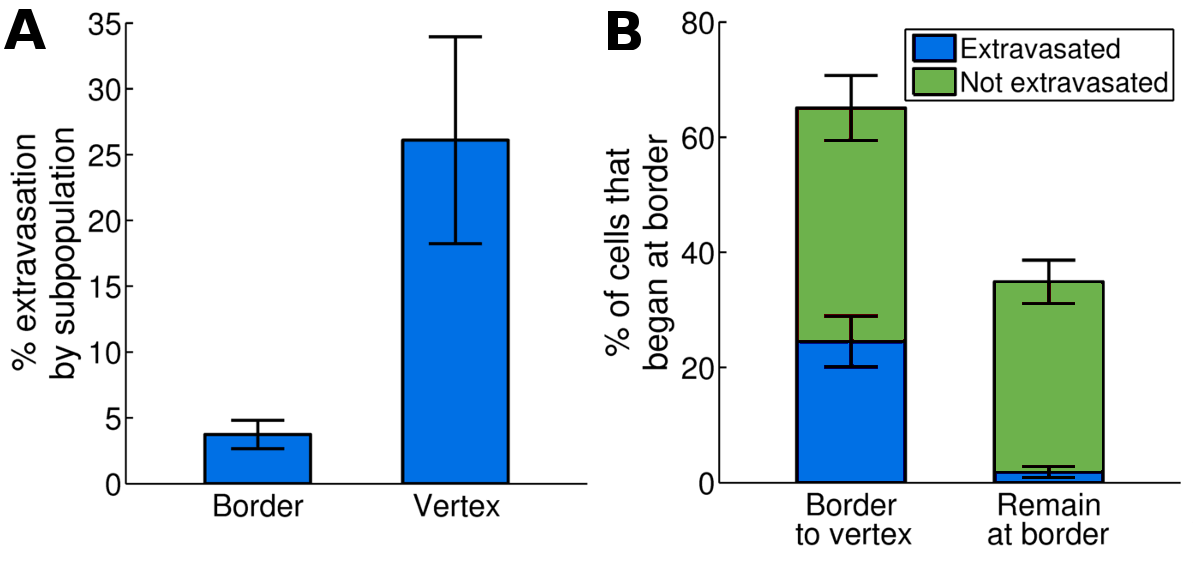}
\caption{\textbf{Cancer cells preferentially extravasate at vertices, even when they first attached to endothelial cell borders.} A: Extravasation of cancer cells in dependence of the location where they transmigrate. B: Extravasation of cancer cells that initially attached to endothelial cell borders, in dependence on whether they extravasate or not and on where they extravasate. {\color{black} Error bars show the standard deviation obtained from 6 (in A) or 4 (in B) repeats of each experiment, performed in separate devices, where in each device about 100-150 extravasating cells were imaged.}}\label{fig:cancerExtravasation}
\end{figure}

	To demonstrate that the geometry of the gap opening dynamics is physiologically relevant, we quantified the characteristics of extravasating cancer cells through monolayers of HUVECs, as shown in Movie \ref{SI:mov:extravasation}. Here, a tumor cell is seen transmigrating through an endothelial monolayer at a tricellular junction as delineated by VE-cadherin GFP, followed by gap-closure after the tumor cell has completely cleared the barrier. Fig. \ref{fig:cancerExtravasation}A shows the ratio of tumor cells that extravasated at vertices, relative to borders. We see that tumor cells preferentially extravasate at the vertices, in line with the previously observed increased frequency of gaps opening there (Fig. \ref{fig:vertexVersusEdgeGaps}G) and similar observations of extravasating neutrophils \cite{burns1997neutrophil}. Moreover, even if cancer cells initially arrest at the border between two endothelial cells, they are much more likely to extravasate at a vertex at later points in time, rather than at the border where they initially attached to, perhaps first through migration on the surface of the endothelium and subsequent preferential attachment to points of exposed basement membrane as a result of inherent EC junctional dynamics (Fig. \ref{fig:cancerExtravasation}B). This could suggest that in addition to the possibility of cancer cells actively signaling to open gaps in the endothelium, endothelial barrier dynamics itself can also present the cancer cells with opportunities to begin the transmigration process.

\section*{Discussion}

The computational model presented in this paper allowed us to study how gaps in an endothelial monolayer initially open, grow, stabilize and finally close, and we identified which physical properties dominantly regulate each stage. 

{\color{black} 
The model simulates a cell monolayer in two dimensions. Adhesion between cells is simulated through binding or unbinding of adhesion complexes located on adjacent cells. These adhesion complexes are dynamically engaging and disengaging as the myosin generated forces cause cell deformations. Because of cell-cell adhesion rupture, gaps between the cells are formed. To perform our simulations, the model is based on a number of assumptions that simplified the model. First of all, due to the typically small height (about $3\mu m$) of endothelial cells \cite{SATO2000127}, we neglected the third dimension perpendicular to the monolayer. However, disruptions of the spatio-temporal dynamics of adhesion molecules and cytoskeletal organization in the third dimensions are likely to impact gap formation. Incorporating such effects into our model would consequently require a 3D model of a cell with more detailed descriptions of the subcellular mechanics. However, the purpose of our model was to demonstrate the broad impact of subcellular mechanical structures on gap formation. For this reason, we modeled the cells in two dimensions and included only radial stress fibers and contractile actin fibers parallel to the membrane. This was motivated by experiments that indicated different roles of these actin structures on gap formation \cite{Huveneers2012}. Each discrete adhesion complex is simulated as one cluster to simulate recruitment of proteins such as vinculin or talin, without increasing the total number of components in the simulation. Myosin generated forces included in the model are assumed to occur only in the direction of the stress fibers or membrane.}

\begin{figure}[!ht]
{\color{black}
	\centering
	\includegraphics[width=0.85\linewidth]{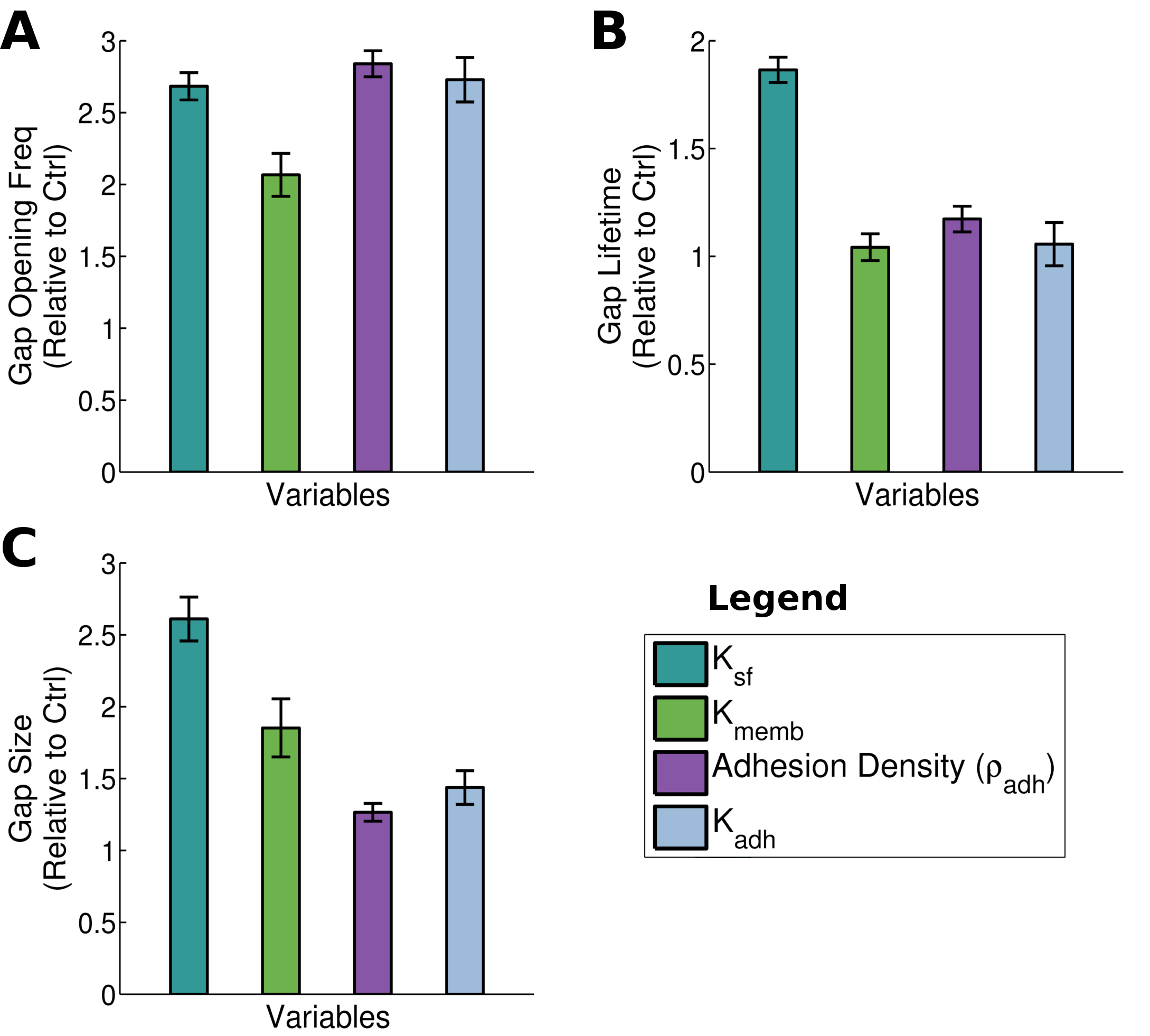}
	\caption{{\bf Interplay of adhesion and cell mechanical properties controls different aspects of gap opening dynamics.} Error bars show to the standard error. All parameters have been reduced one order of magnitude ($x10^{-1}$).  (A) Gap opening frequency. (B) Average lifetime of the gaps. (C) Average size of the gaps.}\label{SI:fig:extreme}
    }
\end{figure}

Supplementary Fig. \ref{SI:fig:extreme} summarizes some of our key conclusions: By comparing our reference case with extreme variations of very low stress fiber or membrane stiffness, {\color{black} we see that both passive mechanical properties and adhesion complex properties are important in controlling gap opening frequency (Supplementary Fig. \ref{SI:fig:extreme}A).} On the other hand, the lifetime and especially size of gaps increases significantly with lower stress fiber or membrane stiffness, since the softer cells are more likely to deform and adapt in response to the opened gap (Supplementary Fig. \ref{SI:fig:extreme}B,C). We also verify that stress fiber stiffness influence is stronger than membrane stiffness influence. In contrast, properties of the cell-cell adhesions strongly affect the frequency of the gap openings, but less so their lifetime or size. Indeed, decreasing the density of adhesion bonds or the adhesion stiffness strongly increase the frequency of forming gaps (Supplementary Fig. \ref{SI:fig:extreme}A), while only marginally affecting the size or lifetime of the gaps (Supplementary Fig. \ref{SI:fig:extreme}B,C). These data thus summarizes our biological model where adhesion properties control the initial formation of gaps, while cell mechanical properties are critical in limiting the size and duration of opened gaps.
	
Our results that gaps open more frequently at vertices than at borders were true over wide ranges of parameters (Fig. \ref{fig:parameterImpact}B,E,H). Only extremely small bending stiffness led to similar frequencies of gaps at vertices and at borders (Fig. \ref{fig:parameterImpact}B). These results also show that earlier experimental observations, where neutrophils were found to extravasate preferentially at endothelial cell vertices, \cite{burns1997neutrophil}, can be explained through the mechanical dynamics of the endothelial monolayer alone. Consequently, this may be a general mechanism for extravasating cells, and we found a similar behavior with extravasating cancer cells (Fig. \ref{fig:cancerExtravasation}). This finding is complementary to the extensive literature that suggests that chemical or mechanical signaling of extravasating immune or cancer cells to the endothelium facilitates extravasation \cite{vestweber2015leukocytes,Reymond2013,Yeh18122017}. There are many potential hypotheses why both the autonomous dynamics of the endothelial monolayer and the bidirectional signaling with the extravasating cells may play a role during extravasation: It may be that initial autonomously forming gaps are important for extravasating cells to sense a gap and they consequently signal to widen the gap or to keep it open. {\color{black} The gap sizes that the model predicts are of the order of magnitude of a few microns, which is enough for extravasating cells to protrude through the gap. Our previous study indicates that tumor cells can squeeze significantly when transmigrating through artificial gaps \cite{CAO20161541}, so the autonomous gaps may be of sufficient size for complete transmigration. Nevertheless, endothelial gaps may widen during transmigration, so crosstalk between the transmigrating cell and the endothelium likely remains an important factor contributing to the likelihood and speed of extravasation.} Then, whether bidirectional signaling or autonomous gap formation dominates the process may be cell type specific. For instance, it is still a major research question why certain cancer cells preferentially metastasize to certain organs \cite{OBENAUF201576}. We may speculate that not only the signaling of the specific primary tumor cells with an organ-specific type of endothelial cells influences the likelihood of extravasation \cite{Reymond2013}. Also, the mechanical properties of the endothelium of the target organs will likely play a major role. Our flexible modeling framework was tested with a HUVECs monolayer, yet, by changing the physical parameters of the model, it may be quickly adapted to other endothelia.

Besides testing our model with different endothelial cells, some other important steps towards validating our model conclusions in vivo will be to test our model with more realistic three dimensional microvasculature with blood flow, embedded in extracellular matrix and surrounded by supporting cells such as pericytes, fibroblasts or, for brain, astrocytes \cite{OBENAUF201576,vestweber2015leukocytes}. Such real, in vivo microvasculature consists of vessels that are curved and exposed to shear stresses due to the flow. That, in turn, may be affected by extravasating cells that may obstruct blood flow. Similarly, matrix stiffness was shown to affect endothelial monolayer integrity \cite{Krishnan2011,AndresenEguiluz2017}.
Some complications in validating our results in vivo involve the lack of available in vitro cultures that are required to provide high throughput, microscopy resolution and level of experimental control that is lacking in vivo, making direct comparison of computational models to in vivo experiments unfeasible. However, the recent rapid progress in developing more complex and organ specific in vitro assays of 3D microvasculature will make such validations feasible in the near future \cite{jeon2015human,chen2017chip,osaki2018engineered}.

{\color{black} 
Our model is based on a number of simplifications. We do not consider the effect of extracellular matrix and substrate stiffness properties on monolayer integrity, despite the known effect of these properties on cell mechanics. It is important to remark that cells on glass may behave very differently than in vivo endothelial vessels. Our model also does not include  the effect of fluid pressure or tangential stress due to fluid flow. Pressure and blood flow would induce additional forces over the monolayer that could affect gap generation processes. For example, it was observed \cite{martinelli2014probing} that tangential flow could induce the strengthening of cell-cell junctions, therefore reducing paracellular extravasation. 

Modeling such complex environments presents a great challenge to both in vitro and in silico models. It is therefore essential to justify assumptions that can reduce this complexity and make the model development feasible. Here, we have assumed that the mechanics of the inside of a cell is determined by a fixed number of stress fibers, although it is known that inside the cell there are different polymer structures such as microtubules and intermediate filaments. Moreover, actin filaments are not fixed in time but appear and disappear depending on their stability and polymerization rates. To simulate all of this with high accuracy would require a completely different model in which the computational cost that would exceed current capabilities. For the purpose of this project, we focused on incorporating essential cell mechanical structures that have been implicated in the regulation of gap formation, and modeled a fixed number of stress fiber similar to other works \cite{Jamali2010, Pathak2016}. Similarly, we have simulated adhesion complexes as discrete elements that can bind two membrane points of neighboring cells. In real cells, adhesion complexes between cells are formed by a great variety of proteins such as VE-cadherins, $\alpha$-catenin, talin or vinculin. While the spatio-temporal dynamics of each of these adhesion molecules likely influences gap formation, no computational model can currently explain their precise organization in adhesion complexes and their resulting effect on gap formation. Consequently, our model included an effective term that describes the force dependent recruitment of adhesions, as observed in different experimental studies \cite{Huveneers2012,oldenburg2014mechanical}.
}

Moreover, there are also challenges to the mathematical modeling of complex 3D microvasculature. Modeling of epithelial sheets in 3D has proved challenging, with some recent interesting progress after decades of mainly focusing on epithelial monolayers in 2D \cite{du2014computational,okuda2015three,alt2017vertex}. These models are based on frameworks such as vertex models, where the dynamics of each cell is incorporated into the dynamics of vertices between cells. {\color{black} There are many other modeling frameworks that can capture different aspects of the complex cell behavior, such as cell based models \cite{TAMULONIS2011217}, immersed boundary models \cite{Rejniak2007} or subcellular element models \cite{Sandersius2008,Sandersius2011}. These modeling frameworks are, however, not directly suitable to predict the formation of gaps at either vertices or borders.} Given these challenges, is was paramount to establish a 2D mathematical model of an endothelial monolayer that was validated with novel experiments and that was able to lead to insights into the mechanisms of endothelial gap formation.

\subsection*{Methods}
\subsection*{Generation of HUVEC monolayers and image acquisition}
		Human umbilical chord vein cells (HUVECs) were transduced with VE-cadherin-GFP using methods described previously \cite{chen2017chip}.
		HUVECs at P7-10 were seeded onto 35 mm glass bottom Mattek dishes (at $3\times 10^5$ cells/dish), which had been plasma treated for 30 seconds previously. Cells were allowed to grow to confluence (beyond 100\%) in EGM-2MV (Lonza) for 3 days before imaging. Dishes were imaged on an Olympus FV1000 confocal microscope with magnifications of 30X  (oil immersion), under an environmental chamber set at 37C and 5\% CO2. The chamber was equilibrated for $\sim ~30$ min prior to the start of image acquisition. For time-lapse videos of junctional dynamics, z-stacks of $40  \mu m$ ($4  \mu m$ steps) were taken at intervals of 3 minutes. 
\subsection*{Analysis of junctional disruption dynamics}
		Time-lapse images were appended and analyzed manually on ImageJ. A single unique junctional disruption is defined as a vertex or border with an observed gap of greater or equal than $3 \mu m$, and are preceded and proceeded at some point in time with a closure (no visible gap in fluorescence greater than $0.6 \mu m$). The number of junctional disruption events was counted for each border and vertex of an image over a total time period of 2 hours. Vertices and borders belonging to the same cells were still considered to be unique. 
\subsection*{Analysis of tumor cell extravasation}
	Tumor cells were suspended in EGM-2MV (Lonza) and a concentration of 15,000 cells/mL, and 1mL of the suspension was gently added to each HUVEC monolayer. Cells were allowed to settle first for $\sim10$ minutes before acquisition of $t=0$ images. For quantification of extravasation, z-stacks were taken at $3 \mu m$ steps at an endpoint of 6 hours to image the entirety of the tumor cell and endothelial monolayer. Any tumor cell that has breached the endothelial layer as evidenced by protrusion extension across and beneath the endothelial layer was considered as ``extravasated''. Delineation of the endothelial barrier is visualized via CD31 staining (Biolegend, Cat \# 303103) for 30 min in EGM-2MV at 37C and 5\% CO2 prior to imaging.

\beginsupplement


\section*{\LARGE\bf Balance of Mechanical Forces Drives Endothelial Gap Formation and May Facilitate Cancer and Immune-Cell Extravasation}
\subsection*{\bf Jorge Escribano, Michelle B. Chen,Emad Moeendarbary,Xuan Cao, Vivek Shenoy, Jose Manuel Garcia-Aznar, Roger D. Kamm, Fabian Spill}

\section*{Supporting Information, Text}
\section*{Detailed derivation of the model}
	
\begin{figure}[h!]
\centering
	\includegraphics[width=0.98\linewidth]{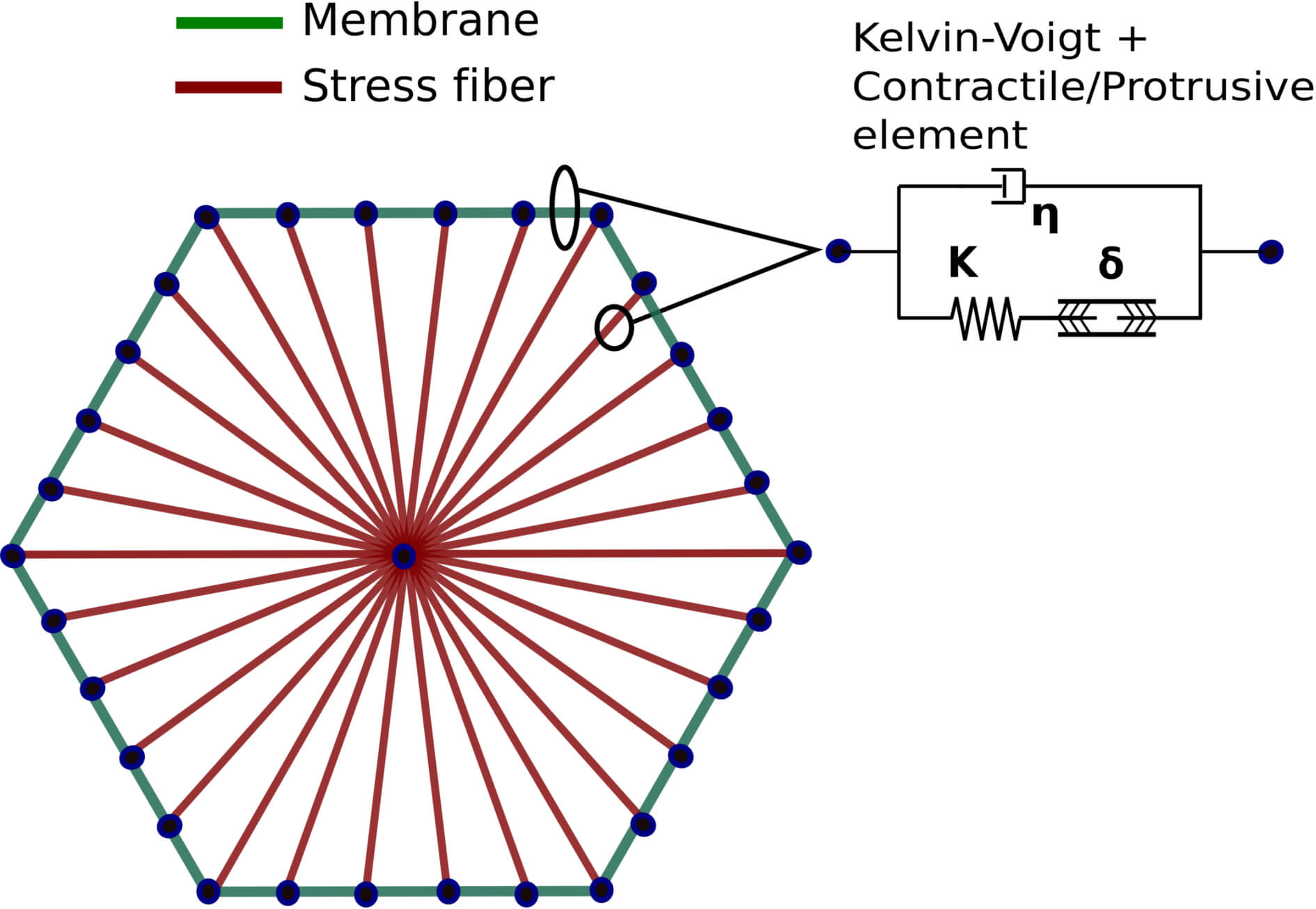}
	\caption{\textbf{Mechanical model of a single cell.} The cell is presented in an initially hexagonal form, divided into a discrete number of membrane points. Physically, our membrane elements connecting the nodes represent the combined lipid bilayer with the actin cortex. Moreover, the nodes are connected to the center by stress fiber structure. Both of them are described by Kelvin-Voigt models with a contractile/protrusive element, but both have different parameters.}
		\label{SI:fig:BasicCell}       
\end{figure}

	The endothelial monolayer is modeled through a number of cells in two dimensions that are connected through cell-cell adhesions. Each cell contains a number of radial contractile actin stress fibers, modeled by viscoelastic springs. Then, viscoelastic springs also represent the combined cell membrane with the adjacent cortex. For simplicity, in the paper we will refer to these elements as membrane elements. Therefore,the whole cell is discretized into nodes, which represent the fundamental degrees of freedom of the resulting network of stress fibers, membrane elements and cell-cell junctions. The cell membrane is discretized into $n_{nodes}$ nodes with a spacing, $l_{n}$ between them. The example of a hexagonal geometry is shown in Fig. \ref{SI:fig:BasicCell}, where all radial stress fibers are connected in the center of the cell. The actual model is independent of cellular geometry. In fact, cell geometry is dynamically changing, as observed in experiments (Movie \ref{SI:mov:ExpReferenceCase}), and that is reflected in the model simulations. Both membrane and stress fibers are considered as viscoelastic elements and are approximated by Kelvin-Voigt structures, in line with previous models \cite{Jamali2010}. We assume that inertial forces have no significant impact on the system, as typical on the cellular scale \cite{Kim2007}, and the forces exhibit inherent randomness due to fluctuations in molecular activities in each cell and due to variability in growth factors or neighboring cells that inevitably affects cell mechanics. The dynamics of each node is then described by a Langevin equation (where inertia effects have been neglected), similar to the one that appeared in other cell mechanical models (see e.g. \cite{Kim2007}):
	
	\begin{equation}
    	\label{eq:1}
	\frac{d\textbf{r}_i}{\textit{d}t}=\frac{1}{\xi}\textbf{F}_{i}
	\end{equation}
	
	Here, $\textbf{r}_i$ corresponds to the current position of each membrane node and cell center $i$, $\xi$ is the medium drag coefficient, and $\textbf{F}_{i}$ represents the sum of all forces due to active, random contractions or through passive mechanical interactions between node $i$ its neighboring nodes:
	\begin{equation}
	\label{eq:2} 
	\textbf{F}_{i}=\textbf{F}^{sf}_{i}+\textbf{F}^{memb}_{i}+\textbf{F}^{adh}_{i}+\textbf{F}^{rep}_{i}+ \textbf{F}^{gen}_{i},
	\end{equation}
	where, $\textbf{F}^{sf}_{i}$ is the force due to radial stress fibers, $\textbf{F}^{memb}_{i}$ is the force due to the membrane and cortex, $\textbf{F}^{rep}_{i}$ is the repulsion force due to contact between different cells, $\textbf{F}^{gen}_{i}$ is the cell generated force due to contractions or protrusions of the stress fiber or membrane elements, respectively. $\textbf{F}^{adh}_{i}$ is the force originating through cell-cell adhesions, represented by VE-cadherin in our stochastic adhesion model described below.
	
\begin{figure}[h!]	
\centering
		\includegraphics[width=0.98\linewidth]{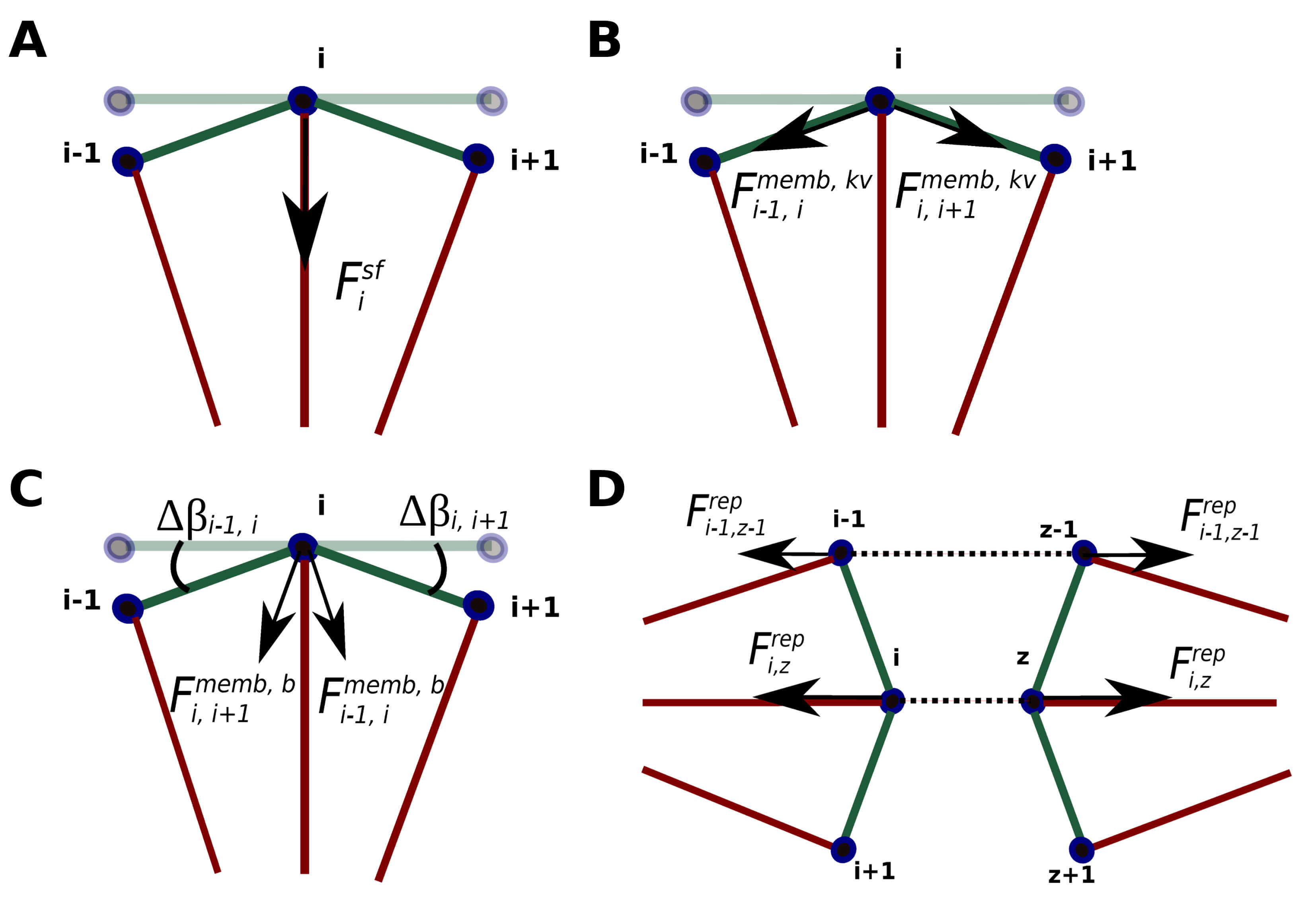}
		\caption{\textbf{Contribution of different passive intracellular forces.} (A) Force due to stress fiber deformations. (B) Force due to membrane in-plane deformation. (C) Force due to membrane bending stiffness. (D) Force due to repulsion between membrane points of different cells.}
		\label{SI:fig:2}       
\end{figure}
	
\subsection*{Model of Passive Intracellular Mechanics}
	
	We now describe all mechanical properties modeled within a single cell. As described before, both radial stress fibers and tangential membrane/cortex segments are modeled with Kelvin-Voigt elements (\ref{SI:fig:BasicCell}). For simplicity, we consider hexagonal cells as a starting point, even though the actual modeling framework is independent of the cell geometry. Stress fibers connect a node in the center of the cell with a node on the membrane. {\color{black} Membrane and center nodes are treated in the same way.} Additional to the Kelvin-Voigt force arising due to deformations in the direction of two membrane points $\textbf{F}^{memb,kv}_{i}$ (Fig. \ref{SI:fig:2}B), the membrane/cortex exhibits bending stiffness resulting in forces due to deformations perpendicular to the membrane, $\textbf{F}^{memb,b}_{i}$ (Fig \ref{SI:fig:2} C). The total force on a membrane node due to deformations of neighboring membrane nodes is thus given by:
	
\begin{equation}
	\textbf{F}^{memb}_{i}=\textbf{F}^{memb,kv}_{i-1,i}+\textbf{F}^{memb,kv}_{i+1,i}+\textbf{F}^{memb,b}_{i-1,i}+\textbf{F}^{memb,b}_{i+1,i}
	\end{equation}
	\begin{equation}
	\vert \textbf{F}^{memb,b}_{i-1,i} \vert = \frac{K_{bend}}{d_{i-1,i}}\cdot(\beta_{i-1,i}-\beta^0_{i-1,i}),
\end{equation}
	where $i-1$ and $i+1$ are, without loss of generality, neighboring points of $i$, $\beta_{i-1,i}$ is the angle denoting deviations from the balance position which is $\beta^0_{i-1,i}$ (see Fig. \ref{SI:fig:2}C) and $d_{i-1,i}$ is the distance between the two neighboring points. For the forces derived from the Kelvin-Voigt structures, $\textbf{F}^{memb,kv}_{i}$ and $\textbf{F}^{sf,kv}_{i}$, the direction corresponds to the vector $\vec{\textbf{k}}$ formed by the corresponding nodes as indicated in the sub-indices ($i-1$ or $i+1$, $i$). For the  bending stiffness component ($\textbf{F}^{memb,b}_{i}$), the direction of the force is perpendicular to $\vec{\textbf{k}}$ (unit vector of the membrane segment), and $K_{bend}$ is the rotational spring constant used to approximate membrane bending effects. 
	
Kelvin-Voigt structures consist of a parallel arrangement of an elastic spring and a viscous damper, with the force  
	\begin{equation}
	F(t)=EA\varepsilon(t)+\frac{d\varepsilon(t)}{\textit{d}t}A \eta.
	\end{equation}
	
	Here, $E$ is the elastic modulus of the spring, $A$ is the area of the cross-section, and $\eta$ is the viscosity of the material (membrane or stress fiber, respectively).
	Forces are implemented for both stress fibers and membrane structures as follows:
	\begin{equation} 
	\textbf{F}^{sf}_{i}= [K_{sf}(l_{n}-l^0_{n})-\eta_{sf}v_{n}]\cdot\vec{\textbf{k}}, 
	\end{equation}
	\begin{equation} 
	\textbf{F}^{memb, kv}_{i}= [K_{memb}(l_{n}-l^0_{n})-\eta_{memb}v_{n}]\cdot\vec{\textbf{k}},  
	\end{equation}
	where $K_{sf}$ and $K_{memb}$ are the stiffness and $\eta_{sf}$ and $\eta_{memb}$ are the drag coefficients of stress fibers and membrane, respectively. $n$ denotes the bar in the network that connects the points $i$ and $i-1$, and could correspond to a stress fiber or a membrane segment. $l_{n}$ is the current length of the element, and $l^0_{n}$ is the rest length. $v_{n}$ corresponds to the velocity at which the element is varying its length, and $\vec{\textbf{k}}$ is the unit vector in the direction of the element.
	
\subsection*{Model of cell-cell junctions}	\label{sec:2.1}

	Endothelial cells are mechanically coupled to neighboring cells through cell-cell adhesions. VE-Cadherin is the major protein in endothelial cell adherens junctions and is known to cluster on the membrane \cite{Huveneers2012}. It is a homophilic protein binding to other VE-cadherins on neighboring cells, and also links cell-cell adhesions to the cytoskeleton \cite{Panorchan2006,dorland2017cell}. Cell-cell adhesions in endothelial cells are very complex, and also include tight junctions. In our model, the precise molecular composition and regulation of the junctions is not relevant. However, it is important to note that several molecular bonds in adhesion complexes are force-sensitive. For instance, a force-dependence of VE-cadherin/VE-cadherin bonds at high forces was known for a long time \cite{Panorchan2006}, and the same paper showed that these bonds have a longer lifetime than the related E-cadherin or N-cadherin bonds present in other cell types. More recent evidence showed that cadherin bonds may also exhibit a catch-bond nature \cite{Manibog2014}, causing the bond-lifetime to initially increase with force. Likewise, the bonds connecting cadherins to the cytoskeleton, for instance through $\alpha$-catenin, were recently found to be described by a catch bond \cite{buckley2014minimal}. 

Our model is thus designed to capture the force dependence of the cell-cell adhesions. Each discretized membrane point can act as a local adhesion cluster that may bind to a membrane point on adjacent cells. If binding occurs, the cells are physically connected through a linear spring with force
	\begin{equation}
	|\textbf{F}^{adh}_{i}| = |\textbf{F}^{adh}_{z}| = K_{adh,p} \cdot( d_{i,z}-L_{adh}^{0}).
	\end{equation}
	Here, $d_{i,z}$ is the distance between the node $i$ on the cell under consideration, and $z$ denotes the node on the adjacent cell. $K_{adh}$ is the stiffness constant of the adhesion complex and $L^0_{adh}$ is the adhesion equilibrium length. ~The direction of the force corresponds to the vector formed by the two points of the adhesion, $i$ and $z$. From now on, we refer to the adhesion complex that binds points $i$ and $z$ with the subindex $p$.

{\color{black}The probability of binding of two membrane points is determined by a rate that depends on the distance between these two points:}
	
\begin{equation}
    k_{bind,p}=
\left\{
\begin{array}{ll}
k_{on}^0\cdot \rho_{adh}\cdot(1-\frac{d_{i,z}}{L_{bind}^{limit}}) & \mbox{if } L_{bind}^{limit}\leq 0,\\\\
0                &  \mbox{if } L_{bind}^{limit} > 0,
\end{array}
\right. 
	\label{eq::bind}
\end{equation}
	
	Here, $k_{on}^0$ it is a binding rate constant and $L_{bind}^{limit}$ is the maximal distance at which two membrane points of neighboring points could bind, and $\rho_{adh}$ is the density of adhesion molecules available for binding.  
	
 To describe the experimentally observed effect that adhesion complexes may strengthen due to clustering of molecules such as VE-cadherin and recruitment of molecules such as talin or vinculin, which themselves have force-dependent binding rates that may lead to positive feedback loops, \cite{Huveneers2012}, our model includes a force dependence of the adhesion complex density and consequently its mechanical properties. Indeed, it was shown that forces play an active role during the strengthening of cell-cell adhesions \cite{Huveneers2012,oldenburg2014mechanical}. Our model thus incorporates a mechanism to reinforce cell-cell junctions in response to forces. This is done by describing the bond density $n_{reinf}$ through a stochastic, force dependent model. For simplicity, and to effectively describe the important case of $n_{reinf} = 0$ that corresponds to complete rupture of the adhesion complex under consideration, we employ a discrete model where $n_{reinf}$ takes on discrete values between $0$ and $n_{reinf}^{max}$. $n_{reinf}^{max}$ corresponds to maximal saturation of the adhesion complex, i.e. maximal binding strength. Each adhesion complex may thus act as a molecular clutch, unbinding for zero density, and fully engaging at $n_{reinf}^{max}$. The resulting stiffness of the spring then changes in a linear way:

\begin{equation}\label{eq:adhesionStiffness}
	K_{adh,p}=n_{reinf,p} \cdot K_{adh}^{0}
\end{equation}

Here, $K_{adh}^{0}$ is the stiffness per unit bond density.
	
	Once a connection between two neighboring cells is formed, (following eq. \ref{eq::bind}), the bond density of the adhesion complex can vary stochastically following a force dependent law where the reinforcement rate is given by:

\begin{equation}\label{eq:reinforcement}
    k_{reinf,p}=
    \left\{
\begin{array}{l}
k_{reinf}^0 \cdot \rho_{adh} \cdot (1- (\lambda_{reinf}-F_{p}^{adh})/\lambda_{reinf}) =   \\\\
k_{reinf}^0 \cdot \rho_{adh} \cdot F_{p}^{adh}/\lambda_{reinf}   \ \ \   \mbox{if } F_{p}^{adh}\leq F_{reinf}^{limit},\\\\
0    \ \ \ \ \ \ \ \ \ \ \ \ \ \ \ \ \ \ \ \ \ \ \ \ \ \ \ \ \ \ \ \    \mbox{if } F_{p}^{adh} > F_{reinf}^{limit}.
\end{array}
\right. 
\end{equation}

    Here, $k_{reinf}^0$ is the binding rate constant for the reinforcement and $\lambda_{reinf}$ is a force constant for shifting the reinforcement curve. Furthermore, $F_{reinf}^{limit}$ is a threshold above which we stop applying the reinforcement. This threshold is set to avoid numerical instabilities due to very large binding rates. It has no physical consequences for the model behavior as long as it is numerically set to much larger values than the force corresponding to the peak lifetime of the catch bond, see Eq. \eqref{eq:catchbondunbinding} and Fig. \ref{SI:fig:CatchSlipExplan} below. This is because for such high forces, the cell-cell adhesion clusters are already certain to have unbound due to the catch-bond nature we are now discussing.
	
Unbinding of single bonds is modeled as a catch bond law \cite{Novikova2013}:
\begin{equation}
	k_{ub,p}=k_{c}^{0} \cdot exp(\theta_c-\theta)+k_{s}^{0}\cdot exp(\theta-\theta_s),
	\label{eq:catchbondunbinding} 
\end{equation}
	
	where $\theta=|F_{p}^{adh}|/F^{0}$ and $\theta_c$, $\theta_s$ are the parameters of the catch and slip bond regimes respectively. $F^0$ is used to normalize the force, $|F_{p}^{adh}|$ is the modulus of the current force on the specific bond and $k_{c}^{0}$ and $k_{s}^{0}$ are the {\color{black} unbinding rate coefficient} for the catch and slip curve respectively.
	
	Since the mechanics of our monolayer is described through connected springs, where the dynamics is exclusively calculated through the forces on the nodes, such springs could hypothetically overlap. A repulsion force on membrane nodes is thus included to prevent different cells from overlapping. This force occurs when two membrane points of two different cells are within a certain small distance range ($L_{rep}$). The magnitude of this force grows with the distance between two membrane points, $i$ and $z$, of the two adjacent cells(Fig. \ref{SI:fig:2}D):
	\begin{equation}
	\textbf{F}^{rep}_{i,z}=K_{rep} \cdot (L_{rep}-d_{i,z}) \cdot \vec{\textbf{j}}
	\end{equation}
	
	Here, $K_{rep}$ is a constant parameter, $d_{i,z}$ is the distance between the two membrane points of different cell, and the direction of the force is obtained as in Fig. \ref{SI:fig:2}D. $L_{rep} $ is the maximum distance at which repulsion is applied and $\vec{\textbf{j}}$ is the unit vector in the opposite direction to the straight line that binds both points ($i$ and $z$).
	
\subsection*{Cell-generated forces}

	 Forces are generated within the cell due to motor activity and cytoskeletal remodeling. Myosin generated forces ($\textbf{F}^{myo}_{i}$) act on both the stress fibers and the cortex membrane in a contractile manner, and protrusive forces ($\textbf{F}^{prot}_{i}$) generated by actin polymerization may lead to forces directed towards the outside of a cell:
\begin{equation}
	   	\textbf{F}^{gen}_{i}=\textbf{F}^{myo}_{i}+\textbf{F}^{prot}_{i}
\end{equation}
	   	
\begin{figure}[h!]
		\centering
		\includegraphics[width=0.98\linewidth]{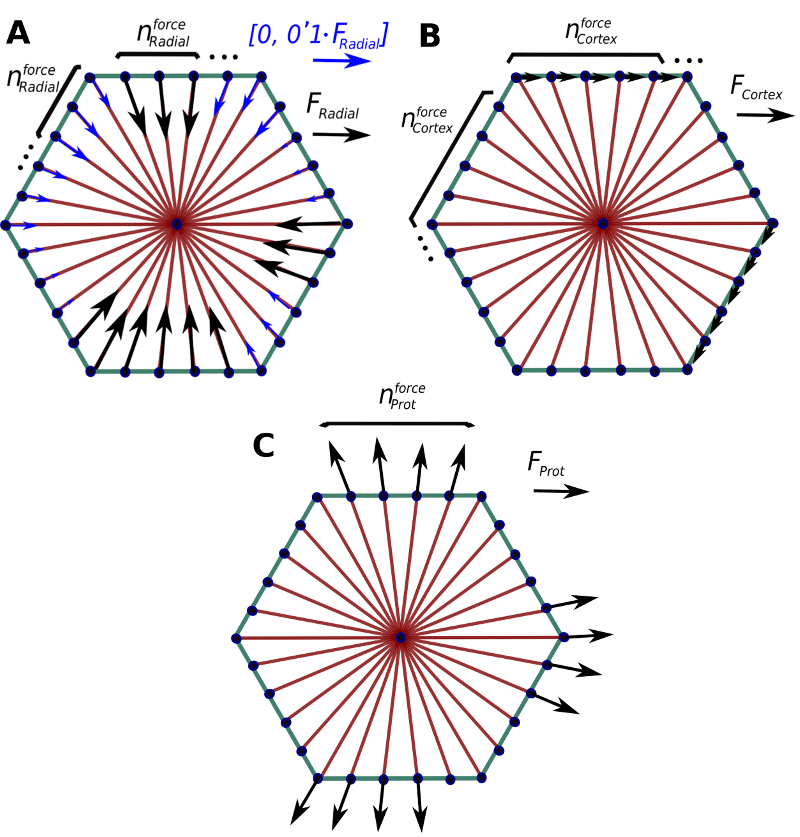}
		\caption{\textbf{Cell generated forces.} (A) and (B) Correspond to myosin forces: Radial force and Cortex force respectively. (C) Protrusive forces.}
		\label{SI:fig:force_exp}  
\end{figure}

	Myosin forces are the result of the combination of two types of forces. The first one, which is generated by the myosin activity in the stress fibers, thus typically results in radial forces. The second one are tangential forces that occur due to contractions of the cortical actin filaments and are directed parallel to the membrane. Both forces have a magnitude of $F_{Radial}^{max}$ and $F_{Cortex}^{max}$ for each stress fiber and membrane segment respectively. Also, both type of forces are not homogeneously distributed throughout all the stress fibers and membrane segments of the cell and are not acting during the whole simulation time. The spatial distribution of the forces is controlled by  $n^{Force}_{Radial}$ and $n^{Force}_{Cortex}$, which represents the number of consecutive stress fibers or membrane segments respectively that have the same force. Each one of this set of segments has a probability for activating the force of $p^{Force}_{Radial}$ and $p^{Force}_{Cortex}$. Depending on this probability, the force for each segment of the different sets is either $F_{Radial}^{max}$, or a baseline contraction, which we simulate to be a random number between [0, $0.1 \cdot F_{Radial}^{max}$] for the radial force case. For the cortex force, depending on the outcome of our Monte-Carlo simulation, the magnitude of the force is either $F_{Cortex}^{max}$ or zero. This probability is calculated for each of the stress fiber and membrane sets at every given time interval indicated by $t^{Force}_{Radial}$ and $t^{Force}_{Cortex}$. These values indicate the time steps when the force activations are recalculated in the Monte-Carlo simulation (see Fig. \ref{SI:fig:force_exp}A,B), and they are much larger than the overall time step used to simulate the whole system (Table \ref{tab:1}). If, following the previous explanation, there is a change in the force due to a random recalculation of the forces for either stress fiber or membrane segment, the force does not change abruptly in one time step. Instead, the force magnitude changes linearly in time from the previous value to the new one in a given total time, $t^{Force}_{Transition}$. This is to mimic the behavior of cells while external conditions remain approximately constant, so no rapid changes in the mechanics within each cell occur (compare for Movie \ref{SI:mov:ExpReferenceCase}).	In this way, cell forces are not homogeneously distributed in time and space, but also do not change abruptly in the absence of external stimuli. External stimuli, for instance, vasoactive agents like thrombin increase intracellular levels of Ca$^{2+}$ and lead to myosin activation \cite{Valent2016a}. This induces changes in traction forces, leading to heterogeneous force distribution which causes the formation of inter-cellular gaps.
	
	Protrusive forces are caused due to actin polymerization at the edges of a cell. {\color{black} For simplicity, protrusive forces are modeled in the same way as contractile forces, and are only distinguished from contractile forces due to the direction of force and the characteristic parameters.} Thus, they act typically in the opposite direction of radial contractile forces (i.e. outward of the cell) and with their own parameters characterizing the typical magnitudes ($n^{Force}_{Prot}$, $p^{Force}_{Prot}$ and $t^{Force}_{Prot}$) (see Fig. \ref{SI:fig:force_exp}C).

\subsection*{Actin remodeling}
	
	The protrusions due to actin polymerisation, as described above, may change the length of stress fibers. {\color{black} Likewise, shrinkage of existing fibers may occur due to depolymerization, or due to severing or buckling and subsequent breakage of fibers \cite{Staiger269}. Other mechanisms leading to changes of the rest length of stress fibers include the addition of sarcomeric units in the middle of stress fibers in response to tension \cite{CHAPIN20122082}. Our model effectively incorporates the dynamical remodeling of stress fibers due to adaptation to the applied forces. For simplicity, we do not consider total depolymerization of a fiber or de novo polymerization of new fibers in response to nucleation. Moreover, we assume that the total amount of F-actin is conserved in a given simulation, i.e. the G-actin available after depolymerization is assumed to quickly polymerize in other fibers.
	
We describe the remodeling of the stress fibers through a change in the rest length of the spring in the Kelvin-Voigt element. This way, we do not explicitly take into account the precise origin of the change in rest length (e.g. whether it is due to actin polymerization, depolymerization or inclusion of sarcomeric units). Stress fibers dynamically remodel by adapting their rest length to their current length at a certain velocity:}
	
\begin{equation}
	\dot{L_{s}^{0}}=v_{s}^{remodel}=K_{remodel} \cdot (L_{s}-L_{s}^{0}) \\
	\label{eq:remodel} 
\end{equation}
	
	Here, $s$ is the index of the stress fiber, $L_{s}$ is the current length of the stress fiber, $ L_{s}^0 $ is the current balance rest length of the stress fiber and $ K_{remodel} $ is a constant describing the rate of length adaptation.
	
\begin{figure}[h!]
		\centering
		\includegraphics[width=0.98\linewidth]{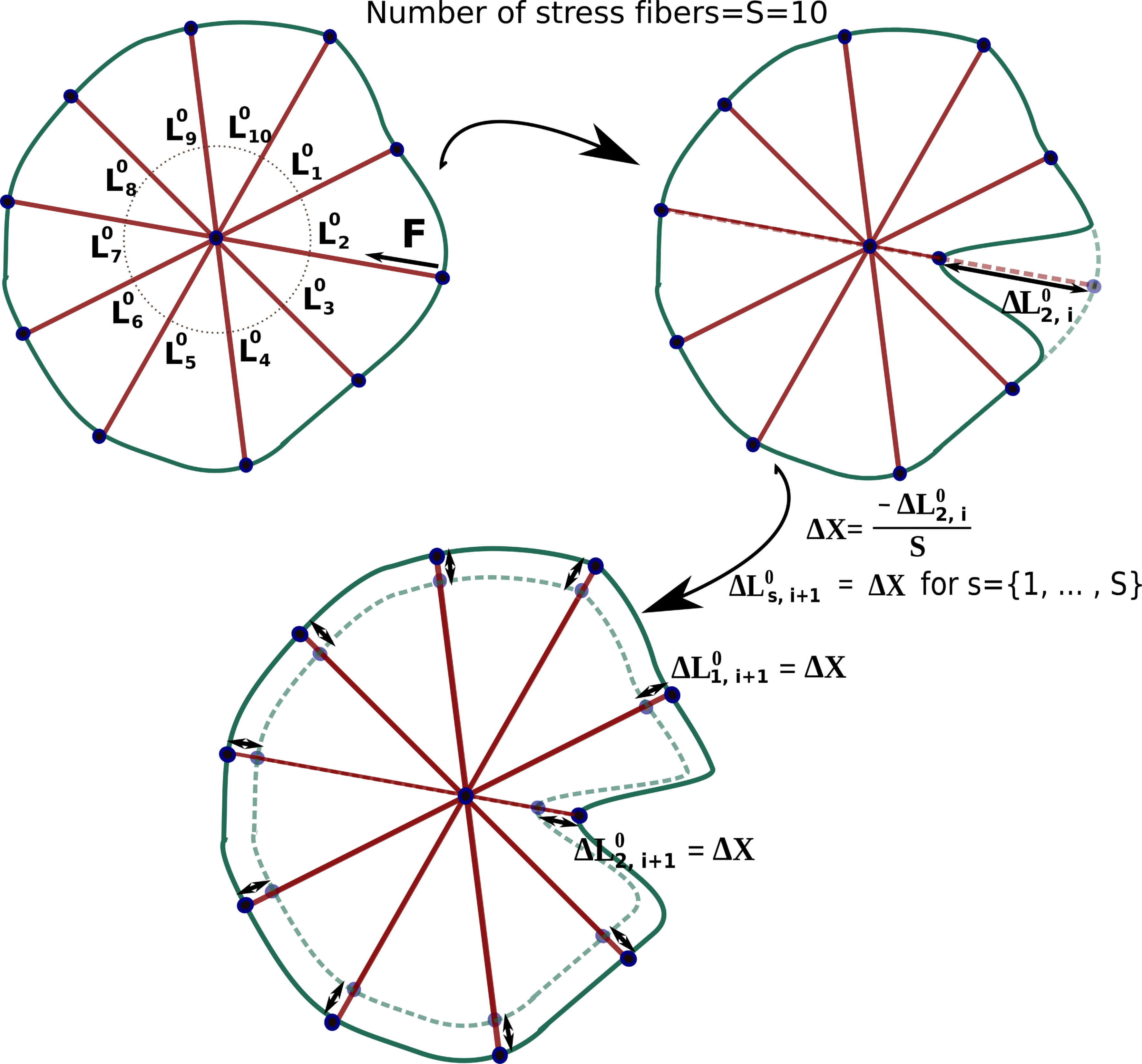}
		\caption{{\bf Stress fiber remodeling.} Due to myosin contractility, a change in the rest length of the stress fiber occurs accordingly to Eq. \eqref{eq:remodel}. This change in rest length is compensated by all the stress fibers in a proportional way. Note that only the rest lengths and not the current length of a stress fiber is modified. 
		}\label{SI:fig:Remodel}
\end{figure}

Since we assume total F-actin conservation, this means that under constant cross-sectional area the total balance rest length of the stress fibers is constant: 
\begin{equation}
	\sum_{s=1}^{S}L_{s}^{0}=const 
	\label{eq:remodelConservation} 
\end{equation}
	Here, $S$ is the total number stress fibers in a cell. For simplicity, we ignore the spatial variations of actin regulators and assume each stress fiber has a similar amount of free barbed ends that polymerize. In order to satisfy Eq. \eqref{eq:remodelConservation}, the rest length of all the stress fibers in a cell is thus modified in the same way (see Fig. \ref{SI:fig:Remodel}).  
		
\section*{Implementation of the model and simulations}
	\label{sec3}
	
	\subsection*{Initial and boundary conditions}
\begin{figure}[h!]
		\centering
		\includegraphics[width=0.98\linewidth]{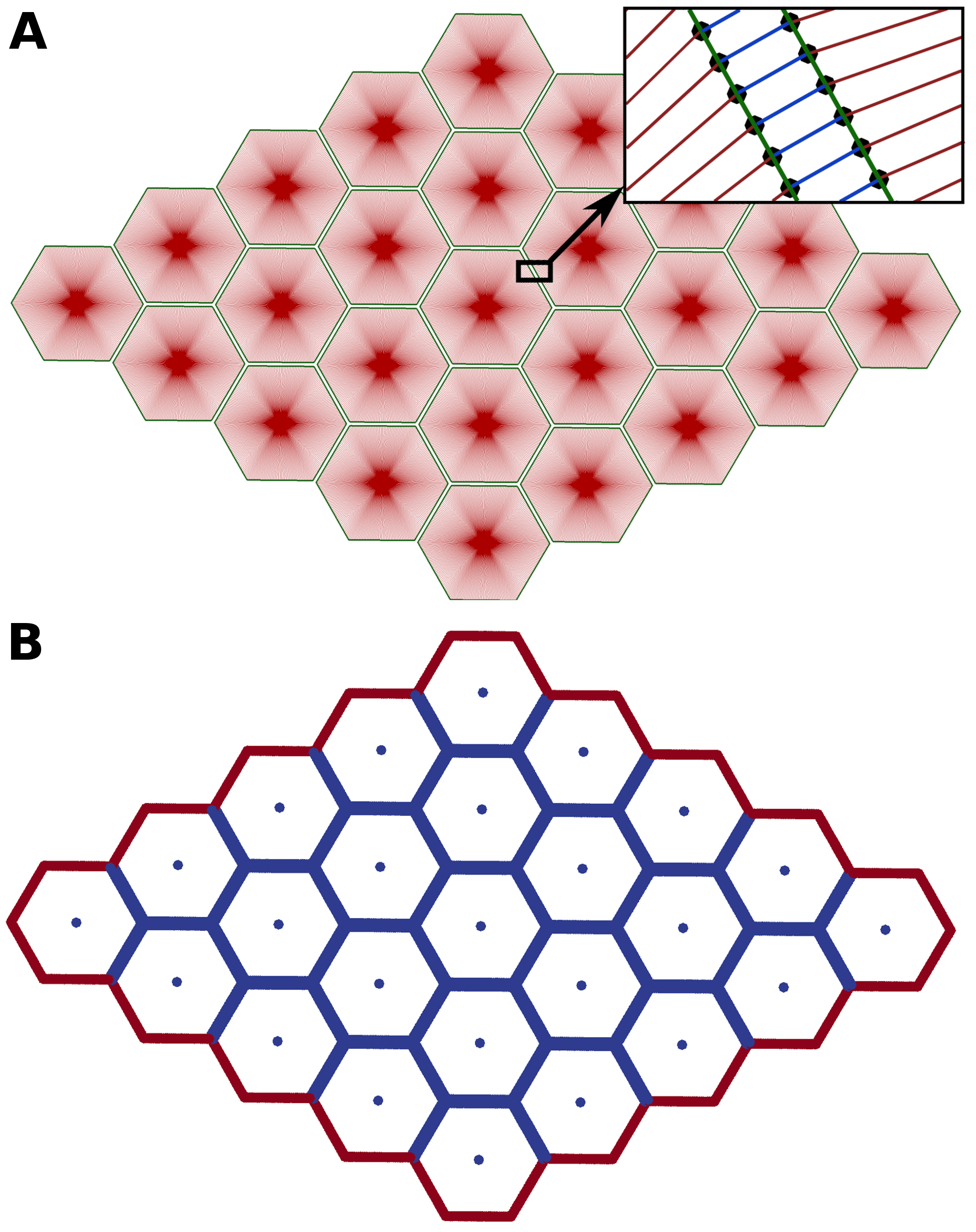}
		\caption{\textbf{Model of the endothelial monolayer.} (A) Cells with a hexagonal shape are in a rest state and fully bound to their neighboring cells. Cell membrane (green), stress fibers (red), cadherin complexes (blue), membrane points (black). (B) Boundary conditions: Points in the boundary of the monolayer (red) are fixed. In blue are membrane points and the cell centers. }
		\label{SI:fig:monolayer}  
\end{figure}

	Fig. \ref{SI:fig:monolayer} shows an example of a cell monolayer, composed of initially hexagonal cells. Here, the monolayer is initialized such that all neighboring adhesions are bound. Membrane points at the edge of the monolayer are encastred, so that the total domain of the monolayer is fixed. This is to mimic our experimental conditions where the cells are placed in fixed devices. The aim of the simulations is to study the mechanisms behind gap formation. To avoid effects due to the boundary conditions, our gap quantification is performed for the cell in the center of the monolayer only. The initial conditions are such that all spring elements are at balance. When the simulation starts, myosin forces and protrusive forces activate and disturb the balance.
	
\subsection*{Quantification of gap formation}
\begin{figure}[h!]
			\centering
			\includegraphics[width=0.98\linewidth]{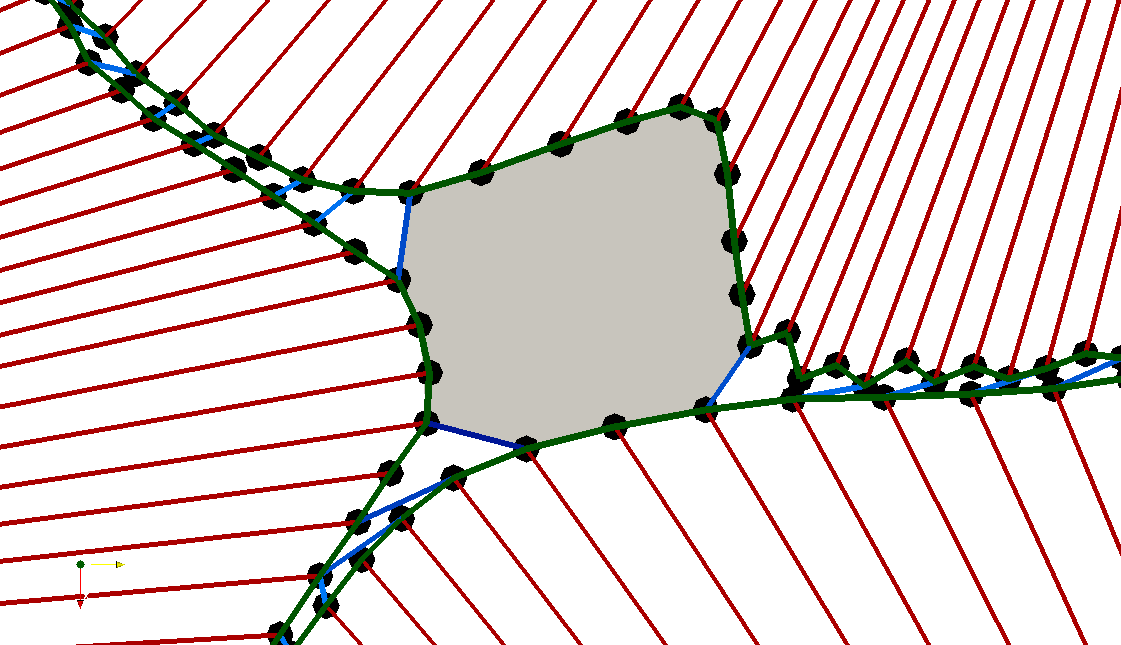}
			\caption{\textbf{Paracellular gap.} A gap (grey area) is delimited by the cell membrane (green) and the adhesion bonds binding the cells (blue). Red: cell stress fibers. Black dots: Membrane points}
			\label{SI:fig:gap}  
\end{figure}

	Gaps are formed between two or more cells as a consequence of the adhesion bond rupture. In the model, as described above, two cells are connected at two adjacent nodes through adhesion complexes that are characterized through a (for simplicity assumed discrete) number of bonds $n_{reinf}$. If $n_{reinf} = 0$, the adhesion complexes of the adjacent cells unbind. However, the unbinding of a single pair of adhesion complexes on adjacent cells does not necessarily imply that the endothelial barrier is breached at that location, as this requires a sufficiently high number of close-by adhesion complexes to rupture. We thus quantify the breached area in between two or more cells that resulted from ruptured bonds and only quantify the rupture events as the formation of a proper gap if the area exceeds a threshold area $A_{GAP,F}$. Likewise, when the gap area drops below a threshold $A_{GAP,C}$, we consider that gap to be closed. $A_{GAP,C}$ is typically chosen to be slightly smaller than $A_{GAP,F}$, since otherwise randomness in the simulation may lead to fluctuations around this detection threshold and thus incorrectly predict gaps to form and close constantly. We differentiate between gaps that are formed at a two cell border and gaps that are formed in the vertex of the cells touching three or more cells. A typical gap and how it is quantified is shown in Fig. \ref{SI:fig:gap}.
	
\subsection*{Implementation}
The model has been implemented in a custom-made C++ program. We used the Eigen3 library for operation with the vectors and matrices used to solve the equations. For the rest of the program we have used standard C++ libraries. {\color{black}The code is available on Github, https://github.com/Escribs/Endothelial-monlayer.}

\subsection*{Simulations}
	
In each simulation, we first set up the initial and boundary condition and then calculate a loop consisting of repeatedly executing the following major steps:
	\begin{enumerate}
		\item Binding and reinforcement of adhesions: First we analyze the position of membrane points of different cells and calculate the probabilities for binding. We perform a Monte-Carlo simulation to see if a new union is formed. We also analyze the reinforcement of adhesions that are already bound.
		\item Node displacement: We analyze the force balance in each node $i$ of the monolayer and calculate the movement of each node with the Langevin equation \eqref{eq:1}. The forces considered are the ones outlined in Eq. \eqref{eq:2}. {\color{black} To avoid instabilities, we set a maximum node displacement. If the node displacement exceeds this threshold, the time step is dynamically reduced when integrating the Langevin equation; a process we perform iteratively. In Fig. \ref{SI:fig:timeStep}, we show that the default time step we fixed is sufficiently low so that simulation results are not significantly affected by this time step.}
		\item Actin Remodeling: Based on the new positions of the nodes, we simulate the remodeling of actin.
		\item Unbinding of adhesion bonds: We update the forces on adhesions after the displacement of nodes, and preform a Monte-Carlo simulation to check if bonds unbind according to Eq. \eqref{eq:catchbondunbinding}. 
		\item Gap formation: Finally we quantify the formation of new and closure of existing gaps and the size of all currently existing gaps.
	\end{enumerate}

\subsection*{Parameter Justification}

\begin{table*}[!ht]
		\begin{tabular}{llll}
			\hline\noalign{\smallskip}
			Parameter& Symbol& Value & Source\\
			\noalign{\smallskip}\hline\noalign{\smallskip}
			Medium drag coefficient & $\xi$ & $4.1 \cdot 10^{-3}$ ($kg/s$)& \cite{Pathak2016} \\
			Membrane stiffness & $K_{memb}$ & $2.5 \cdot 10^{-3}$  ($kg/s^2$)& \cite{Dorn2016}\\
			Stress fiber stiffness & $K_{sf}$ &  $1.25 \cdot 10^{-4}$ ($kg/s^2$) & \cite{Pathak2016}\\
			Rotational spring constant & $K_{bend}$ & $7.5 \cdot 10^{-17}$ ($Nm$)& \cite{Dorn2016} \\
			Membrane viscosity & $\eta_{memb}$ & $1.109 \cdot 10^{-3}$ ($kg/s$)& \cite{Dorn2016}\\
			Stress fiber viscosity & $\eta_{sf}$ & $1.109 \cdot 10^{-3}$ ($kg/s$)& \cite{Dorn2016}\\
			Force to normalize parameters in unbinding law & $F^{0}$ & 0.008 ($nN$)& Adjusted from \cite{Panorchan2006}\\
			Non-dimensionalized force of catch curve & $\theta_{c}$ & 0.01 & Adjusted from \cite{Panorchan2006}\\
			in unbinding law &  & \\
			Non-dimensionalized force of slip curve & $\theta_{s}$ & 4& Adjusted from \cite{Panorchan2006} \\
			in unbinding law &  & \\
            {\color{black} Unbinding rate coefficient} for catch curve& $k_{c}^{0}$ & 0.27 $s^{-1}$ & Adjusted from  \cite{Panorchan2006} \\
            {\color{black} Unbinding rate coefficient} for slip curve& $k_{s}^{0}$ & 0.27 $s^{-1}$ & Adjusted from  \cite{Panorchan2006}  \\
			Binding rate for adhesions & $k_{on}^{0}$ & $15.3$ ($ \mu m^2/(mol \cdot s)$) & Estimated from\\ 
            at maximum distance &  & &  unbinding law\\ 
			Binding rate for adhesion reinforcement & $k_{reinf}^{0}$ & $11.5$ ($ \mu m^2/(mol \cdot s)$) & Estimated from \\ 
            at zero force & & & unbinding law \\
			Adhesion complex density	&  $\rho_{adh}$ &  $21$ $(mol/\mu m^2)$ &   \cite{Chen2016a}  \\	
			Limit distance for cadherin binding & $L_{bind}^{limit}$ & 0.95 ($\mu m$)& Estimated \\
			Force constant for reinforcement curve	& $\lambda_{reinf}$ &  $10$ $nN)$ &  Adjusted from   \\
            & & & unbinding law\\
			Force threshold to stop applying reinforcement	&  $F_{reinf}^{limit}$ &  $0.06$ $(nN)$ &  Adjusted  from \\
            & & & unbinding law\\
			Adhesion complex stiffness constant per bond & $K_{adh}^{0}$ & $2 \cdot 10^{-4}$ ($kg/s$)& Estimated \\
			Adhesion complex equilibrium length & $L_{adh}^{0}$ & 0.1 ($\mu m$)&  \cite{Chtcheglova2010} \\
			Maximum number of cadherins per clutch & $n_{c}^{max}$ & 8 & Estimated\\
			Maximum force due to radial contraction & $F_{Radial}$ & 0.775 (nN)& Adjusted from \cite{Rabodzey2008}\\
			Maximum force due to cortical tension & $F_{cortex}$ & 0.025 (nN)& \cite{Hassinger2017}\\
			Maximum force due to protrusion & $F_{Prot}$ & 0.08 (nN)& Estimated\\
			Force recalculation time for radial force& $t^{Force}_{Radial}$ &  25 $min$ & Estimated\\
			Force recalculation time for cortical force& $t^{Force}_{Cortex}$ &  25 $min$ & Estimated\\
			Force recalculation time for protrusive force& $t^{Force}_{Prot}$ &  25 $min$ & Estimated\\
			Force transition time & $t^{Force}_{Transition}$ &  2 $min$ & Estimated\\
			Number of nodes with similar radial force &  $n^{Force}_{Radial}$ &  5 & Estimated\\
			Number of nodes with similar cortical force &  $n^{Force}_{Cortex}$ &  10 & Estimated\\
			Number of nodes with similar protrusive force &  $n^{Force}_{Prot}$ &  20 & Estimated\\
			Force activation probability for radial force& $\textit{p}^{Force}_{Radial}$ & 0.01 & Estimated \\
			Force activation probability for cortical force& $\textit{p}^{Force}_{Cortex}$ & 0.01 & Estimated \\
			Force activation probability for protrusive force& $\textit{p}^{Force}_{Prot}$ & 0.1 & Estimated \\
			Constant for repulsion & $K_{rep}$ & $ 10^{-3}$ ($kg/s^2$ )& Estimated \\
			Maximum distance to apply repulsion & $L_{rep} $& 0.05 ($\mu m$) & Estimated \\ 
			Remodel rate constant & $k_{remodel}$ & $ 0.025 s^{-1}$&  Estimated\\
			Hexagon side length & $l_{hexagon}$ & $25(\mu m)$& Estimated \\
			Distance between membrane points & $l_{n}$ & $625(nm)$& Estimated\\
			Minimum area for gap formation & $A_{GAP,F}$ & $2(\mu m^{2})$& Estimated \\
			Area for gap closing & $A_{GAP,C}$ & $1.5(\mu m^{2}) $& Estimated\\
			Time step & $\Delta t$ & $ 1.26 (s) $& \\
			

		\end{tabular}
		\caption{Reference model parameters used in the simulation.}
		\label{tab:1}
\end{table*}
	
\begin{figure}[!ht]
{\color{black}
	\centering
	\includegraphics[width=0.98\linewidth]{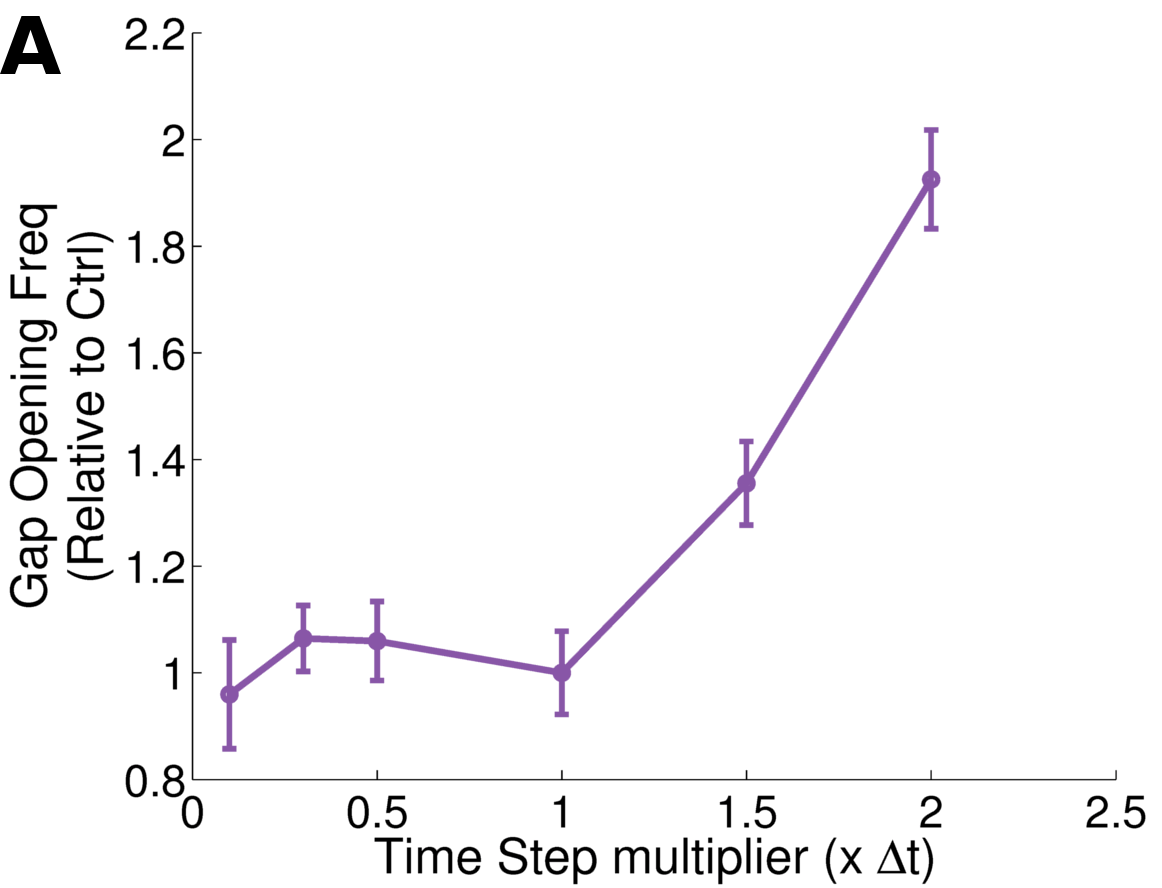}
    \caption{{\bf Time step analysis.} The gap opening frequency depends on the time step used in our numerical simulations. Note that the time step multiplier is relative to the reference case (multiplier $=1$). Error bars are the standard error. The results confirm that the time step selected for the reference case is low enough to ensure convergence of the results.  }\label{SI:fig:timeStep}
    }
\end{figure}

The parameters of the reference case are summarized in Table \ref{tab:1}. Parameters for the unbinding law are adjusted to match data from \cite{Panorchan2006}, where distributions of forces were shown to be required to break single VE-cadherin/VE-cadherin bonds in HUVECs. Binding and binding reinforcement values have been adjusted according to the unbinding rates: At low loading rates binding and unbinding rates are within the same order of magnitude, but one is higher depends on the binding distance. For intermediate loading rates, binding is predominant due to the reinforcement of bonds. For high load rates, bonds ultimately rupture as unbinding is predominant past the catch bond peak of maximal lifetime, estimated from \cite{Panorchan2006}. Radial forces in the monolayer were reported by \cite{Rabodzey2008}. For protrusive forces we have selected values around four times lower than radial forces. Geometrical parameters of the model are estimated based on our experimental images, and the geometrical parameters of adhesions are extracted from \cite{Chtcheglova2010}. Stress fiber stiffness is obtained from models of epithelial cells \cite{Pathak2016}. Membrane stiffness is within the range of values reported for two neighboring membrane ring segments in \cite{Dorn2016}, and similar to measurements of cellular cortex stiffness in endothelial cells \cite{Grimm2014}. Membrane and stress fiber viscosity are within the order of magnitude of values reported for viscous drag coefficients for filament shrinkage \cite{Dorn2016}. The value of the membrane bending stiffness is within two orders of magnitude of values reported in \cite{Dorn2016} by the cell height of approximately $10\mu m$. The medium drag coefficient is also within one order of magnitude of values used in another model for epithelial cell monolayers \cite{Pathak2016}. Typical values reported for cortical tension are of the order of $400 pN/\mu m$ \cite{Sens2015}. If we assume that the membrane has a thickness of $20nm$, the resulting force is around $8 pN$, which is considerably smaller than the active contraction forces. For the density of adhesion molecules, we use VE-cadherin as a proxy, given its established role as critical player to form effective cell-cell adhesions. The numerical value is estimated from similar models focusing on E-cadherin in epithelial cell models \cite{Chen2016a}. {\color{black}The time step is a numerical parameter that is chosen sufficiently low, so that results are stable and convergent. In Fig. \ref{SI:fig:timeStep} we can see that reducing the time step does not change the results significantly. However, as we increase the time step, gap opening frequency is changing. This is because these larger time steps become of the order of magnitude of adhesion binding and unbinding time scales that affect the dynamics of the system. The choice of the time step is consequently an optimal choice that guarantees convergence of our results while being as large as possible for optimal simulation performance.}

{\color{black}To reproduce experimental results, we performed a parameter fitting for those parameters of the model that were not fixed according to published literature values, as outlined above. Generally, our fitting was performed through adjusting published parameter values in similar systems, e.g. epithelial cells. First, we had to adjust the viscosity of the system, together with the cell generated forces and the stiffness of both stress fibers and membrane segments. These parameters combined, control the overall velocity at which the nodes move. This is important to reproduce the velocity at which cells deform and move, and to fit the experimentally observed lifetime of the gaps. Binding and unbinding parameters are very important to reproduce gap opening frequencies, since they control the rupture and binding of the adhesion complexes. The binding law is estimated and we set the values so they are consistent with the unbinding law rate from literature values. Finally, bending stiffness has a strong influence on the location where the gaps are generated (border vs vertex), and was consequently adjusted so that computational results matched our experimental data.

We varied selected parameters within certain ranges in the results section of our main text. We focused on fold changes that have maximal impact on the shown results. Data for physiological ranges that can motivate these fold changes are rarely available. We considered fold changes typically within one order of magnitude from the reference case, which is likely obtainable through experiments or in different physiological conditions. However, even if such fold changes are not physiologically relevant, they demonstrate the principle impact of the represented physical parameters, and therefore, their importance, on gap formation.
}

\clearpage

\clearpage

\begin{figure*}[ht]
		\centering
		\includegraphics[width=0.3\linewidth]{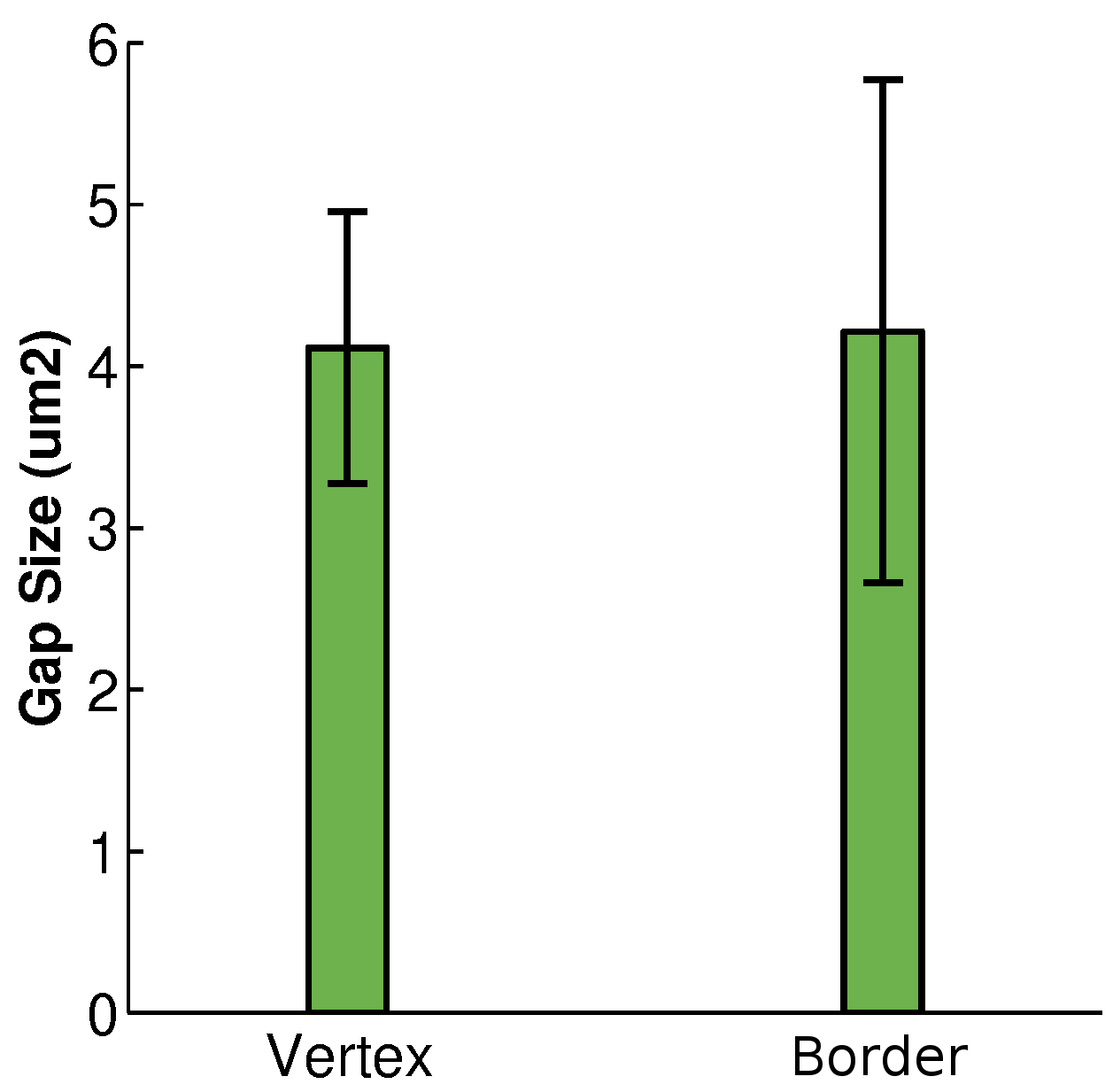}
		\caption{\textbf{Gap sizes predicted from simulations with reference parameters.} Average size of the gaps generated at the vertices and borders. Parameters are the reference values as in Table \ref{tab:1} and error bars correspond to standard deviation of sample=30.}\label{SI:fig:SizeRef}
\end{figure*}

\clearpage

\begin{figure*}[ht]
	\centering
	\includegraphics[width=0.6\linewidth]{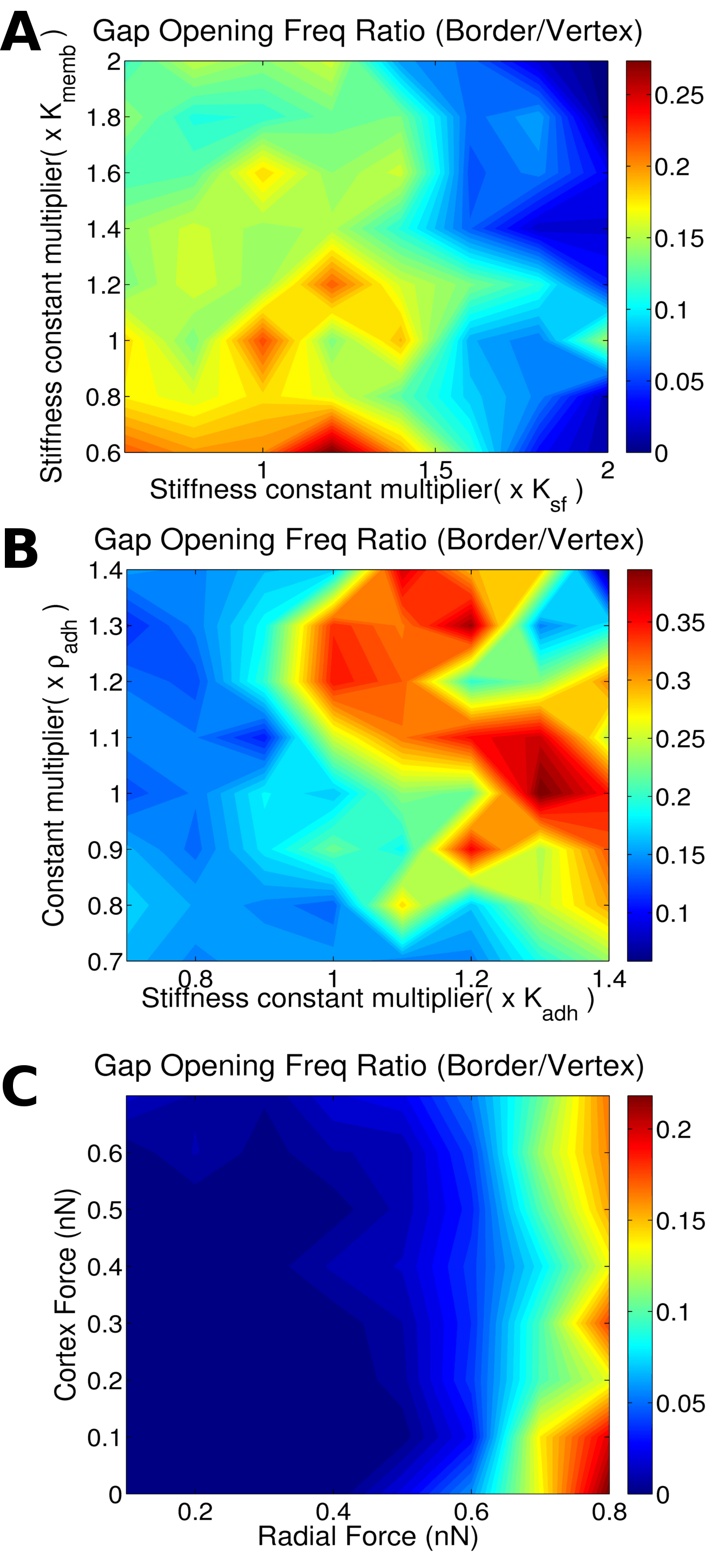}
	\caption{\textbf{Effect of two parameter variation on gap opening location.} Shown is the ratio of gaps that occur at a two cell border divided by the gaps that originate at a three cell vertex. A shows results varying membrane and stress fiber stiffness. B shows properties of cell-cell junction are changed: cadherin stiffness versus cadherin density (binding rate). C shows results for varying cortical and radial force.}\label{fig:parameterImpact2D_gapFreqRatio}
\end{figure*}

\begin{figure*}[ht]
	\centering
	\includegraphics[width=0.8\linewidth]{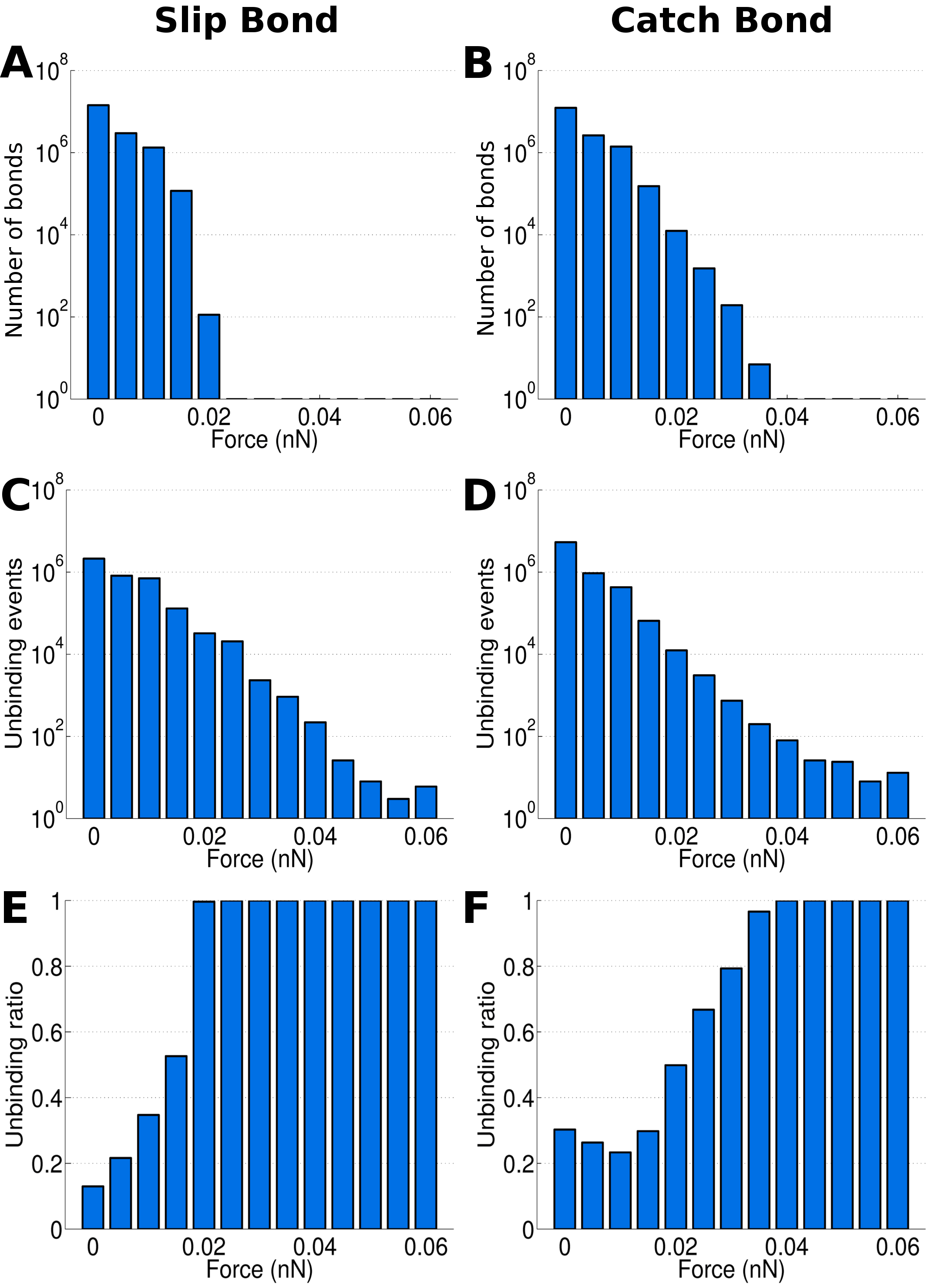}
	\caption{\textbf{Forces on bonds, comparing a pure slip bond and the catch bond law used as reference in the paper.} First row shows force histogram of cadherins that are bound. Second row cadherins force at which cadherins unbind. Third row shows the ratio obtained by dividing unbound cadheins by the sum of unbound cadherins and bound cadherins ($ub/(ub+b)$, where ub and b corresponds to unbound and bound cadherins respectively.)  }\label{SI:fig:CatchSlipHist}
\end{figure*}

\begin{figure*}[ht]
	\centering
	\includegraphics[width=0.8\linewidth]{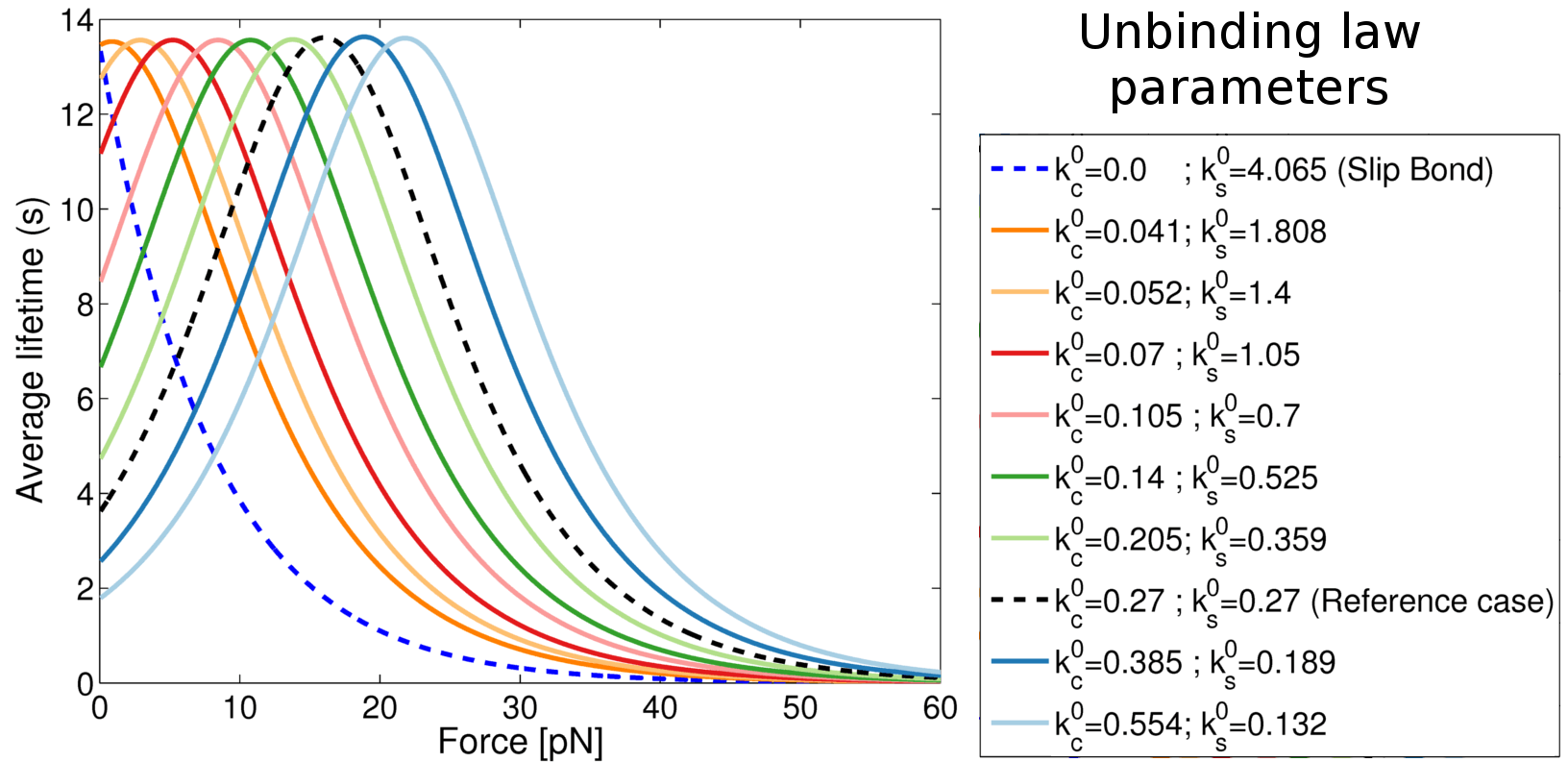}
	\caption{\textbf{Shift of force of maximal catch bond lifetime.} Lifetime average for the bond in dependence on the force for different unbinding laws. Legend shows the parameter variation to obtain the different curves.}\label{SI:fig:CatchSlipExplan}
\end{figure*}

\begin{figure}[!ht]
{\color{black}
	\centering
	\includegraphics[width=0.98\linewidth]{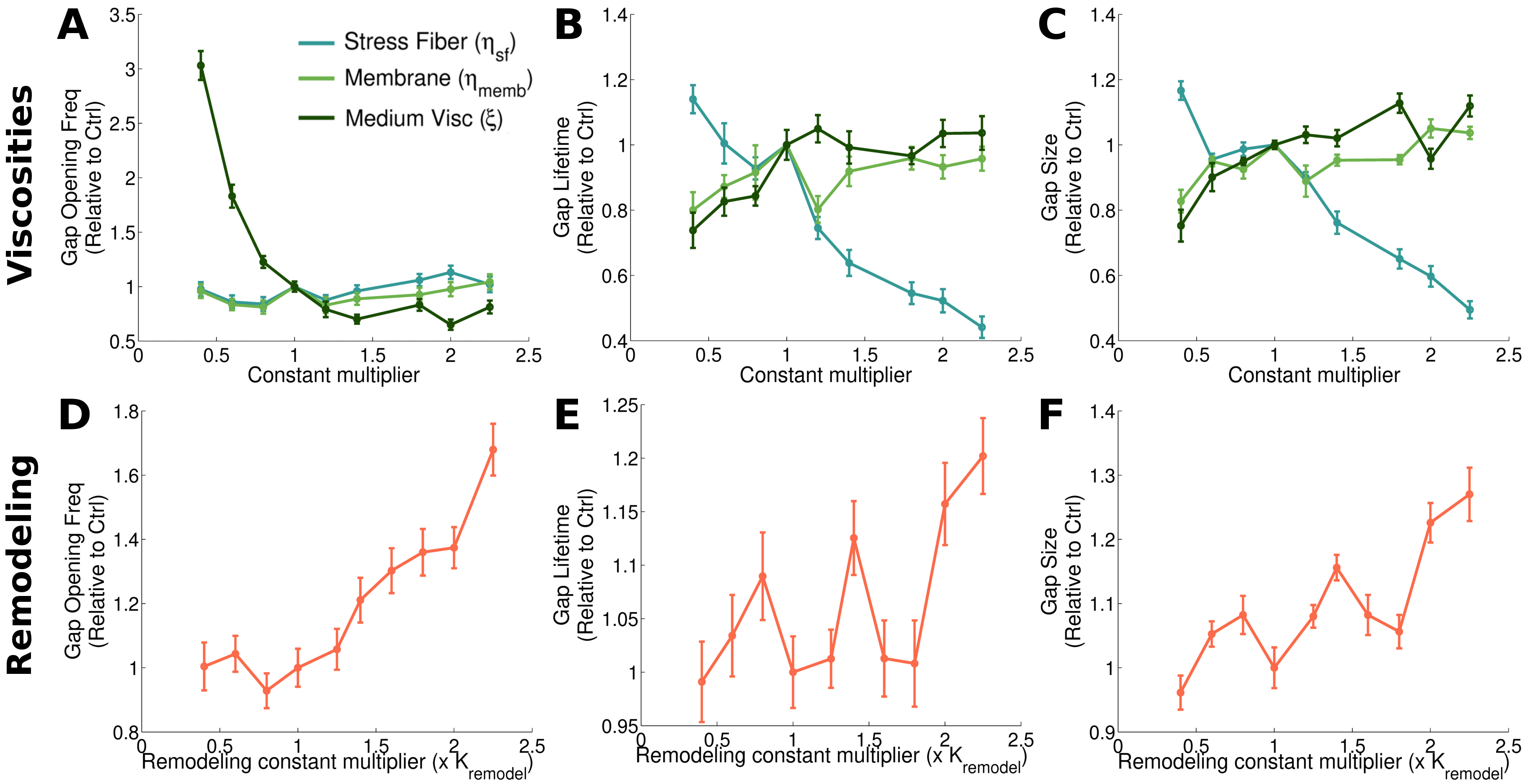}
	\caption{{\bf Effect of the viscosities and the remodeling rate on the gap opening dynamics.} Gap opening frequency, average lifetime and size in each column. Note that the point where x and y coordinates are 1 corresponds to the reference case. Error bars represent standard error. In the first row (A, B, C), results for medium and dashpot viscosities of the stress fibers and membrane and varied. Increasing viscosity reduces node movement, stabilizing monolayer dynamics. Medium viscosity has a higher effect on gap opening dynamics since it affects the overall timescale of all mechanical parts of the model. The stress fiber dashpot strongly influences gap lifetime and size; this is similar to the dominating effect of stress fiber stiffness over membrane stiffness on gap lifetime and size (Fig. \ref{fig:SensLifeSize}A,B) . The second row (D, E, F) shows the effect of varying the constant for remodeling rate. Increasing the remodeling rate implies that cells are able to adapt their permanent shapes faster in response to deformations. Therefore, the frequency of gap openings increases with the remodeling rate (D). The gap lifetime and size broadly also increase, but less strongly then the opening frequency.}\label{SI:fig:viscosity:remodeling}
}
\end{figure}


\beginmoviecaptions
\begin{figure}[ht]
	\centering
	\caption{\textbf{Simulation of the endothelial monolayer dynamics.} Gaps are more likely to appear in the vertex of three cells than at a two cell border. Green denotes the cell membrane, red the inside of a cell, with darker red being the stress fibers. Parameters are the reference values as in Table \ref{tab:1}}\label{SI:mov:ReferenceCase}
\end{figure}

\begin{figure}[ht]
	\centering
	\caption{\textbf{Experimental observation of an endothelial monolayer dynamics.} Dynamics of a monolayer of HUVEC cells, corresponding to Fig. \ref{fig:vertexVersusEdgeGaps}D-F}\label{SI:mov:ExpReferenceCase}
\end{figure}

\begin{figure}[ht]
	\centering
	\caption{\textbf{Stresses on the cell-cell adhesions.} Homogeneous contractions are applied to a hexagonal cell, showing that stresses naturally concentrate on the adhesions at vertices, as opposed to the adhesions at the border. This leads to a faster gap generation at these areas. }\label{SI:mov:stress}
\end{figure}

\begin{figure}[ht]
	\centering
	\caption{\textbf{Altered monolayer dynamics due to low stiffness in the stress fibers.} Not only gap opening frequency is increased under these conditions but also, gaps are critically larger compared to the reference case.}\label{SI:mov:lowKsf}
\end{figure}

\begin{figure}[ht]
	\centering
	\caption{\textbf{Altered monolayer dynamics due to high stiffness in the stress fibers.} Gap opening frequency is strongly suppressed for very stiff stress fibers.}\label{SI:mov:highKsf}
\end{figure}

\begin{figure}[ht]
	\centering
	\caption{\textbf{Altered monolayer dynamics due to low bending stiffness.} Membranes can easily deform when forces are applied, reducing gap formation. }\label{SI:mov:lowKbend}
\end{figure}

\begin{figure}[ht]
	\centering
	\caption{\textbf{Altered monolayer dynamics due to high bending stiffness.} Cells tend to be more rounded, provoking a concentration of stress at the adhesions at the vertices and leading to gap generation in these zones. Gaps are bigger and difficult to close.}\label{SI:mov:highKbend}
\end{figure}

\begin{figure}[ht]
	\centering
	\caption{\textbf{Altered monolayer dynamics due to slip bonds.} Gap opening frequency is clearly increased under these conditions compared to the reference case (based on catch bonds).}\label{SI:mov:slipbond}
\end{figure}

\begin{figure}[ht]
	\centering
	\caption{\textbf{Cancer cell extravasation occurring at endothelial cell vertex.} MDA-MB-231 tdTomato (red) extravasating through a HUVEC endothelial monolayer at a vertex. Endothelial junctions are visualized via VE-cadherin GFP (green). After successful transmigration, cancer cells spreads and migrates below the monolayer, followed by the re-sealing of the endothelial gap. Images are taken every 12 minutes.}\label{SI:mov:extravasation}
\end{figure}

\end{document}